\documentclass[twocolumn]{aastex631}
\usepackage{booktabs}
\usepackage{multirow}
\usepackage{enumitem}
\usepackage{natbib}
\usepackage{subfigure}
\shorttitle{SWASTi-CME}
\shortauthors{Mayank et al.}
\graphicspath{{./}{figures/}}

\begin{document}

\title{SWASTi-CME: A physics-based model to study CME evolution and its interaction with Solar Wind}

\correspondingauthor{Prateek Mayank}
\email{prateekmayank9@gmail.com}

\author[0000-0001-8265-6254]{Prateek Mayank}
\affiliation{Department of Astronomy, Astrophysics and Space Engineering, Indian Institute of Technology Indore, India}

\author[0000-0001-5424-0059]{Bhargav Vaidya}
\affiliation{Department of Astronomy, Astrophysics and Space Engineering, Indian Institute of Technology Indore, India}
\affiliation{Center of Excellence in Space Sciences India, IISER Kolkata, India}

\author[0000-0003-2740-2280]{Wageesh Mishra}
\affiliation{Indian Institute of Astrophysics, Bangalore 560034, India}

\author[0000-0003-2693-5325]{D. Chakrabarty}
\affiliation{Space and Atmospheric Sciences Division, Physical Research Laboratory, Ahmedabad, India}



\begin{abstract}

Coronal mass ejections (CMEs) are primary drivers of space weather and studying their evolution in the inner heliosphere is vital to prepare for a timely response. Solar wind streams, acting as background, influence their propagation in the heliosphere and associated geomagnetic storm activity. This study introduces SWASTi-CME, a newly developed MHD-based CME model integrated into the Space Weather Adaptive SimulaTion (SWASTi) framework. It incorporates a non-magnetized elliptic cone and a magnetized flux rope CME model. To validate the model's performance with in-situ observation at L1, two Carrington rotations were chosen: one during solar maxima with multiple CMEs, and one during solar minima with a single CME. The study also presents a quantitative analysis of CME-solar wind interaction using this model. To account for ambient solar wind effects, two scenarios of different complexity in solar wind conditions were established. The results indicate that ambient conditions can significantly impact some of the CME properties in the inner heliosphere. We found that the drag force on the CME front exhibits a variable nature, resulting in asymmetric deformation of the CME leading edge. Additionally, the study reveals that the impact on the distribution of CME internal pressure primarily occurs during the initial stage, while the CME density distribution is affected throughout its propagation. Moreover, regardless of the ambient conditions, it was observed that after a certain propagation time (t), the CME volume follows a non-fractal power-law expansion ($\propto t^{3.03-3.33}$)  due to the attainment of a balanced state with ambient.

\end{abstract}



\section{Introduction} \label{section1}

Coronal Mass Ejections (CMEs) are gigantic expulsions of magnetized plasma from the Sun, serving as the primary drivers of space weather and causing hazardous effects on ground and space-based technological systems \citep{schwenn_2006_space, webb_2012_coronal}. Therefore, it is essential to understand the eruption of CMEs at their birthplace, their propagation in the corona, and their subsequent evolution in the heliosphere in order to accurately predict their time of arrival and enable timely response. While the dynamic evolution of CMEs near the Sun is governed by the Lorentz force, their further evolution in the inner heliosphere is influenced by the pre-conditioned ambient medium they encounter. Interactions with the ambient solar wind and other large-scale structures, such as CMEs and stream interacting regions, play a significant role in shaping the morphological and dynamical properties of CMEs \citep{bojanvrnak_2010_the, savani_2010_observational, mishra_2017_assessing, geyer_2023_interaction}. Remote-sensing observations provide valuable insights into the kinematics, density, and mass of CMEs, but become increasingly challenging as CMEs become more tenuous in the inner heliosphere \citep{harrison_2012_an, mishra_2013_estimating, temmer_2014_asymmetry}. Additionally, routinely available white-light remote observations lack information about the magnetic field, temperature, and composition. In-situ measurements offer direct measurements of CME speed, density, temperature, magnetic field, and other parameters. However, these in-situ measurements have limited spatial coverage, impeding the acquisition of a complete picture of the spatial and temporal distribution of CME properties. In this scenario, data-driven numerical models can be highly useful in acquiring missing information to complement remote-sensing observations and in-situ measurements. 

In the above context, numerous studies have been conducted to develop inner-heliospheric models that aim to continuously track the evolution of CMEs in the interplanetary medium, considering the limitations of imaging and in-situ observations. These approaches can be broadly classified into two categories: magneto-hydrodynamic (MHD) simulations and analytical modeling. In the category of MHD models, the WSA-ENLIL+Cone model \citep{dusanodstrcil_2004_numerical} has been widely used, which implements a non-magnetized CME with a simple cone geometry \citep{xie_2004_cone}. Another MHD model, SUSANOO-CME \citep{shiota_2016_magnetohydrodynamic}, utilizes a spheromak-type magnetic CME. EEGGL+AWSoM\_R \citep{jin_2017_dataconstrained} combines the Gibson-Low flux rope model \citep{gibson_1998_a} with the AWSoM MHD solar wind model to compute CME plasma properties in the inner-heliosphere. Similarly, EUHFORIA \citep{pomoell_2018_euhforia} also utilizes the MHD approach by incorporating a non-magnetized cone \citep{pomoell_2018_euhforia}, magnetized spheromak \citep{verbeke_2019_the}, and magnetized flux rope CME \citep{maharana_2022_implementation}.

Among the analytical models, the drag-based model \citep{bojanvrnak_2013_propagation} exclusively accounts for the effect of solar wind drag to determine the CME arrival time. This model presumes that the CME evolution is primarily governed by its interaction with ambient solar wind within the inner-heliosphere. Its updated version, the DBEM \citep{matejadumbovi_2018_the}, carries out multiple runs to yield an ensemble of arrival time and speed results by taking into account the uncertainty in input values obtained from coronagraph images. The ELEvoHI \citep{rollett_2016_elevohi} model further incorporates heliospheric imager observations to estimate CME properties, while its upgraded version \citep{jrgenhinterreiter_2021_dragbased} also considers the deformation of the CME front due to ambient solar wind conditions. Likewise, the newly developed ANTEATR-PARADE \citep{kay_2021_modeling} drag-based model accounts for changes in CME size and shape while it propagates in the inner-heliosphere.

By using the above mentioned CME models as well as observations, several studies have demonstrated how the properties of CMEs can be influenced by ambient solar wind conditions as they propagate through the inter-planetary space. Interactions of CMEs with high-speed streams (HSS), stream/corotating interaction regions (SIR/CIR), and the heliospheric current sheet (HCS) are particularly responsible for these changes. These interactions have an impact on the kinematic and structural evolution of CMEs. For example, fast CMEs tend to deflect eastward during their propagation in the inner-heliosphere, and this deflection is further enhanced in the presence of a CIR \citep{heinemann_2019_cmehss, liu_2019_numerical}. Furthermore,  when a CME propagates within an HSS, it experiences a lesser drag force compared to when it encounters the boundary of an HSS \citep{kay_2022_beyond}. Some studies have also reported the pancaking of the CME structure \citep{riley_2004_kinematic, anilraghav_2020_the} and the deformation of the leading edge of CMEs \citep{jrgenhinterreiter_2021_dragbased, sudar_2022_influence}.

In addition, the characteristics of the ambient solar wind have a significant impact on the internal magnetic and thermodynamic properties of CMEs \citep{winslow_2021_the, davies_2022_multispacecraft}. However, only a limited number of numerical studies have quantitatively analyzed the influence of these interaction on the evolution of the internal properties of CMEs \citep{scolini_2021_exploring, verbeke_2022_overexpansion, scolini_2022_causes}. The primary challenge in exploring the 3D properties of CMEs lies in accurately separating the CME structure from the surrounding ambient SW. Due to the dearth of comprehensive quantitative investigations into CME-SW interactions, the extent to which ambient SW conditions can affect CME evolution remains unclear. Moreover, it is uncertain which changes arise from the inherent evolution of the CME and which are influenced by the ambient SW. Furthermore, it is unknown whether all properties of the CME are uniformly affected or if they exhibit distinct responses. 

\defcitealias{mayank_2022_swastisw}{Paper I}
In this study, we have introduced a novel CME model, referred to as SWASTi-CME, which has been integrated into the Space Weather Adaptive Simulation (SWASTi) framework \citep[][referred to as `Paper I' hereafter]{mayank_2022_swastisw}. The SWASTi-CME module encompasses two models: an elliptic cone model and a flux rope model. The elliptic cone model represents a non-magnetic CME with a simplified geometry, while the flux rope model incorporates magnetic properties and captures the main deformations occurring in the coronal region. The primary aim of this research is twofold. Firstly, we aim to validate SWASTi-CME for both single and multiple CME events. Secondly, we investigate the impact of ambient solar wind conditions on the CME evolution.

To accomplish the first objective, we have selected two Carrington rotation (CR) periods: one with multiple halo CMEs occurring near solar maxima, and another with a single halo CME occurring near solar minima. We then compared the simulated CME events corresponding to these CR periods with the in-situ observations at L1 using OMNI data \citep{Papi2020}.

Regarding the second objective, we have established a simulation setup to investigate the impact of ambient solar wind on CME behavior. In this setup, we have employed a passive tracer technique to precisely isolate the three-dimensional structure of CMEs within the heliosphere. By employing this technique, we analyze the alterations in the morphological and dynamical properties of CMEs as they evolve in the inner-heliosphere.

The paper is organized as follows. Section \ref{section2} presents the description of the SWASTi-CME model, encompassing the ambient solar wind, cone and flux rope CME. In Section \ref{section3}, the simulation results are analyzed and compared with in-situ observations at of CMEs at L1, focusing on the case study of CMEs in CR2165 and CR2238. The simulation setup designed to investigate the impact of the ambient solar wind medium on CMEs dynamics is outlined in Section \ref{section4}, followed by the results and analysis presented in Section \ref{section5}. Finally, Section \ref{section6} provides a summary and discussion of the findings.

\section{SWASTi-CME} \label{section2}
SWASTi (Space Weather Adaptive SimulaTion framework) is a newly developed 3D physics-based heliospheric model designed for space weather research and forecasting \citepalias{mayank_2022_swastisw}. This numerical framework consists of a semi-empirical coronal domain (from 1 R$_\odot$ to 21.5 R$_\odot$) and a magnetohydrodynamic (MHD) based heliospheric domain (from 0.1 AU to 2.1 AU). The CME module of SWASTi comprises two models: the cone model and the flux rope model, which are inserted into the self-consistently propagating background solar wind (SW). The specifics of these CME models, as well as the setup for the solar wind, are detailed in the following subsections. 

\subsection{Ambient Solar Wind}
The solar wind module of SWASTi relies on photospheric magnetograms as observational inputs to calculate plasma properties of the solar wind in the inner heliosphere. For this particular study, we utilized zero-point corrected GONG synoptic maps. The solar wind speed at the initial boundary of the MHD domain ($V_{in}$), located at 0.1 AU, is determined using a modified version of the Wang-Sheeley-Arge relation \citep[WSA;][]{arge_2003_improved}:

    \begin{equation}\label{eq:WSA}
       V_{{in}} = v_{1} + \frac{v_{2}}{(1+f_s)^{\frac{2}{9}}} \times \Bigg(1.0 - 0.8\,exp \Bigg(-\bigg(\frac{d}{w}\bigg)^{\beta}\Bigg)\Bigg)^{3}
   \end{equation}

where, $v_{1}, v_{2},$ and $\beta$ are independent parameters whose values are taken to be 250 km s$^{-1}$, 675 km s$^{-1}$ and 1.25, respectively. $f_s$ and $d$ are areal expansion factor of flux tube and minimum angular separation of the foot-point from coronal hole boundary, whereas $w$ is the median of $d$ value for those fieldlines that reaches the location of Earth. The initial density is estimated by considering the conservation of kinetic energy at 0.1 AU, while the temperature is derived assuming a constant thermal pressure. Additionally, the magnetic field is calculated using a velocity-dependent empirical relation. For further details reader may refer to Section 2 of \citetalias{mayank_2022_swastisw}. In this work, the number density and magnetic field associated to fast solar wind ($v_{\rm fsw}$ = 650 km s$^{-1}$) were taken to be $n_{\rm fsw}$ = 200 cm$^{-3}$ and $B_{\rm fsw}$ = 300 nT, respectively, and pressure was kept constant at 6.0 nPa at the initial boundary of MHD domain.

After obtaining all the necessary parameters at 0.1 AU from semi-empirical coronal domain, the time-dependent ideal MHD equations are solved in 3D space using the PLUTO code \citep{mignone_2007_pluto}. A uniform static grid is used in spherical coordinates, with a finite volume method employed for the numerical simulation. The set of conservative equations used in the MHD simulation is described in \citetalias{mayank_2022_swastisw}, with a specific heat ratio of 1.5 for the solar wind plasma. In this work, the heliospheric computational domain extends from 0.1 AU to 2.1 AU in the radial direction (r), $\pm$60\textdegree{} in azimuthal direction ($\theta$) and 0\textdegree{} to 360\textdegree{} in meridional direction ($\phi$), with a grid resolution of $256 \times 61 \times 181$, respectively.

\subsection{Elliptic Cone Model}
SWASTi utilizes the cone model to simulate the inner-heliospheric dynamics of CMEs. The cone model has a simple non-magnetic geometry and has been used in numerous studies for forecasting and assessing the CME properties in the inner-heliosphere \citep[EUHFORIA,][]{pomoell_2018_euhforia, dusanodstrcil_2004_numerical}. A common approach in most of the existing MHD models is to describe the CME structure by a constant angular width, propagation direction and speed (e.g., EUHFORIA, ENLIL). They assume the cross-section of CME to be circular with a varying radius. However, from the observations of CMEs in white-light images of the corona, we know that the geometry of CME front is not always circular rather an ellipse. In this work, we propose a more realistic structure for the cone CME model by incorporating the angular height as well. The proposed method allows for the generation of CMEs with elliptical cross-sections, rather than assuming a circular shape. In this manner, the model's geometry better aligns with the observations of CMEs in the corona and enhances the realism of the simulation.

We use the graduated cylindrical shell \citep[GCS;][]{thernisien_2011_implementation} method to estimate the structural parameters of the CME at 0.1 AU (R$_{\rm in}$). Afterwards, the CME is inserted in the MHD domain as a homogeneous plasma cloud with uniform radial speed ($v_{\rm CME}$), density ($\rho_{\rm CME}$) and temperature ($T_{\rm CME}$). The time dependent injection at the initial boundary (R$_{\rm in}, \theta, \phi$) is governed by the following equations:

\begin{equation} \label{eq:cme_condition}
    \Big(\frac{\phi-\phi_{\rm CME}}{w(t)}\Big)^2 + \Big(\frac{\theta-\theta_{\rm CME}}{{h(t)}}\Big)^2 < 1
\end{equation}
\begin{equation} \label{eq:width_time}
    w(t) = \varphi_{\rm hw} \cdot \mathrm{sin}\Big[\frac{\pi}{2}\frac{(t-t_{\rm onset})}{t_{\rm half}}\Big]
\end{equation}
\begin{equation} \label{eq:height_time}
    h(t) = \varphi_{\rm hh} \cdot \mathrm{sin}\Big[\frac{\pi}{2}\frac{(t-t_{\rm onset})}{t_{\rm half}}\Big]
\end{equation}
\begin{equation} \label{eq:half_time_width}
    t_{\rm half} = \mathrm{R}_{\rm in} \cdot \frac{\mathrm{tan\big(\varphi_{hw}}\big)}{v_{\rm CME}}
\end{equation}

where, $\phi_{\rm CME}$ and $\theta_{\rm CME}$ are the longitude and latitude of the CME center, $w(t)$ and $h(t)$ are the time dependent width and height of the CME, $\varphi_{\rm hw}$ and $\varphi_{\rm hh}$ are angular half-width and half-height of the CME determined from the coronagraph images, $t_{\rm onset}$ is the injection starting time and $t_{\rm half}$ is the half of the insertion time of the CME. Equation \ref{eq:cme_condition} ensures the formation of ellipsoidal structure of CME at the boundary of simulation domain, where the ambient solar wind parameters are replaced with CME values. The angular width and height of the elliptic cone CME are modulated sinusoidally with time at R$_{\rm in}$, with the half-time given by equation \ref{eq:half_time_width}, following a similar approach as \cite{pomoell_2018_euhforia}.

\subsection{Flux Rope Model}

\begin{figure*}
   \centering
   \includegraphics[width = \textwidth]{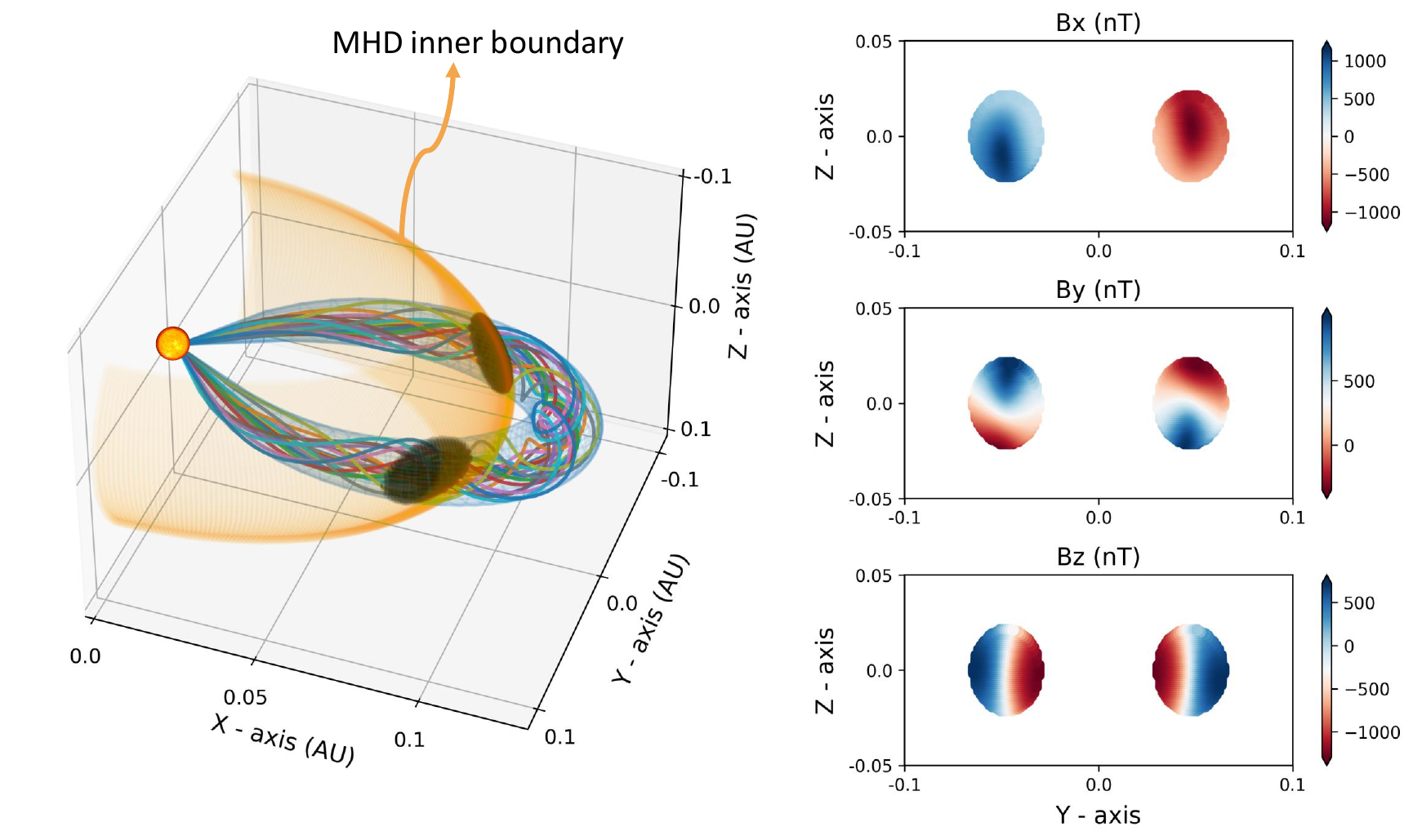}
   \caption{Figure illustrates the flux rope model implemented in SWASTi-CME. The left subplot depicts the evolution of the formed flux rope structure as it propagates radially and crosses the inner boundary of the MHD domain. On the right subplot, the magnetic values are displayed for the overlapping region between the flux rope and the MHD boundary, shown as black circle in the left subplot.}
   \label{fig:Fri3D_insertion}
\end{figure*}

To simulate the flux rope CME, SWASTi employs the Flux Rope in 3D \citep[FRi3D;][]{isavnin_2016_fried} model. FRi3D is an analytical model that incorporates the 3D magnetic field configuration of CME and is capable of considering the major deformations of CME in reconstructing its global geometrical shape.  It has been widely utilized in previous works on CME MHD simulation, such as \cite{maharana_2022_implementation} where they implemented this model in EUHFORIA and compared its performance with the spheromak model, and \cite{singh_2022_ensemble} where they used it to define the geometry of the flux rope CME in their MHD simulation.

In this work, we have used the FRi3D model to construct the 3D magnetized shell of the CME at 0.1 AU, serving as the initial state for the MHD domain. The geometry of the CME forms a classic croissant-like shape that is anchored at both ends to the Sun. The cross-section of CME is assumed to be circular with the radius,  $R(\phi)$, varying in proportion to the heliocentric distance, as described below:

\begin{equation} \label{eq:fri3d_radius}
    R(\phi) = \frac{R_{\rm t}}{R_{\rm p}} \cdot r(\phi)
\end{equation}
\begin{equation} \label{eq:fri3d_hh}
    R_{\rm p} = R_{\rm t} \cdot \mathrm{tan}(\varphi_{\rm hh})
\end{equation}
\begin{equation} \label{eq:fri3d_axis}
    r(\phi) = R_{\rm t} \cdot \mathrm{cos^n}\Big(\frac{\pi}{2} \frac{\phi}{\varphi_{\rm hw}}\Big)
\end{equation}

where $R_{\rm t}$ is the heliospheric distance of apex of the CME axis (toroidal height), $R_{\rm p}$ is the radius of cross-section at the apex (poloidal height) which depends on $R_{\rm t}$ and half-height of CME ($\varphi_{\rm hh}$), and $r(\phi)$ is the radial distance of the CME axis at $\phi$ from the center of the footpoints at the solar surface. This radial distance is defined by the equation \ref{eq:fri3d_axis}, which depends on the half-width of the CME axis ($\varphi_{\rm hw}$) and the parameter $n$ controls the front flattening of the CME structure. See Figure 1 and 2 of \cite{isavnin_2016_fried} for further reference.

The CME structure is populated with magnetic fieldlines that have a low and constant twist, in accordance with the findings of \cite{hu_2015_magnetic}. The magnetic fieldlines are twisted based on equation \ref{eq:fri3d_radius}, and their shape are defined by equation \ref{eq:fri3d_axis}. Meanwhile the strength of magnetic fields follows a normal distribution in the poloidal plane (perpendicular to the flux rope axis) at each axial point. The analytical form of magnetic field strength at a perpendicular distance $\tilde{r}$ from the flux rope axis is following:
\begin{equation} \label{eq:fri3d_magnetic_equation}
    B = B_{\rm axis} \cdot \mathrm{exp}\bigg(\frac{-1}{2\sigma^2}\left[\frac{R_{\rm t} \cdot \tilde{r}}{R_{\rm p} \cdot r(\phi)}\right]^2\bigg)
\end{equation}

where, $B_{\rm axis}$ is the total magnetic field strength on the flux rope axis at $\phi$, which is based on the user defined total flux, conserved along the flux rope cross-section, and $\sigma$ is the standard deviation coefficient of the distribution. Alternatively, other studies have also used other magnetic profiles to magnetize the FRi3D geometry. For example, \cite{singh_2022_ensemble} used an analytic form of a constant-turn magnetic field in a torus shape given by \cite{mvandas_2017_magnetic}. The employed magnetic field profile shown in \ref{eq:fri3d_magnetic_equation}, with default value of sigma as 2 has been further demonstrated in Appendix \ref{Appendix1}.

Figure \ref{fig:Fri3D_insertion} depicts the CME structure with magnetic field modeled using the above described method and its insertion in the MHD domain. Once the FRi3D CME structure is formed, with its leading edge at 0.1 AU, it is allowed to expand radially in a self-similar manner by increasing $R_{\rm t}$, while keeping everything else constant over time. In this manner, the toroidal and poloidal heights of CME ($R_{\rm t}$ and $R_{\rm p}$) increases with time as:

\begin{equation} \label{eq:fr_insertion}
    R_{\rm t}(t) = R_{\rm t}(0) + v_{\rm t}\cdot t
\end{equation}
\begin{equation} \label{eq:fr_insertion2}
    R_{\rm p}(t) = R_{\rm p}(0) + v_{\rm p}\cdot t
\end{equation}

where $R_{\rm t}$(0) and $R_{\rm t}$(0) are the initial toroidal and poloidal heights when the leading edge is at 0.1 AU. $v_{\rm t}$ and $v_{\rm p}$ are the speeds at which toroidal and poloidal heights increase, respectively. The effective speed of CME apex ($v_{\rm CME}$) is the summation of $v_{\rm t}$ and $v_{\rm p}$, as estimated from coronagraphic observations at 0.1 AU. As the leading edge of the CME surpasses 0.1 AU, the initial boundary of the MHD domain is substituted with the values of the overlapping flux rope CME at the intersection region. Similar to cone CME, both density ($\rho_{\rm CME}$) and temperature ($T_{\rm CME}$) are uniform within the flux rope. However, in contrast to cone CME, the local speed inside the flux rope CME varies radially, based on the local toroidal speed of the FRi3D geometry.

The left panel of Figure \ref{fig:Fri3D_insertion} shows the region where the FRi3D CME and MHD domain overlap, indicated by two black circles in the orange plane. The subplots in the right panel of Figure \ref{fig:Fri3D_insertion} display the magnetic field values of this overlapping region. The flux rope CME values corresponding to this intersection region are initialized at the inner-boundary of the MHD domain, and are updated as the leading edge progresses. The insertion process stops when its local speed at boundary matches with the speed of the solar wind stream at the corresponding grid points. Given the variability in solar wind speeds, we use a rounded value of their average for this purpose. This process of leg disconnection ensures that there are no abrupt and significant changes in the pressure gradient, thereby facilitating the smooth insertion of subsequent CMEs.

\section{Case Study of CR2165 \& CR2238} \label{section3}

\begin{figure*}
   \centering
   \includegraphics[width = \textwidth]{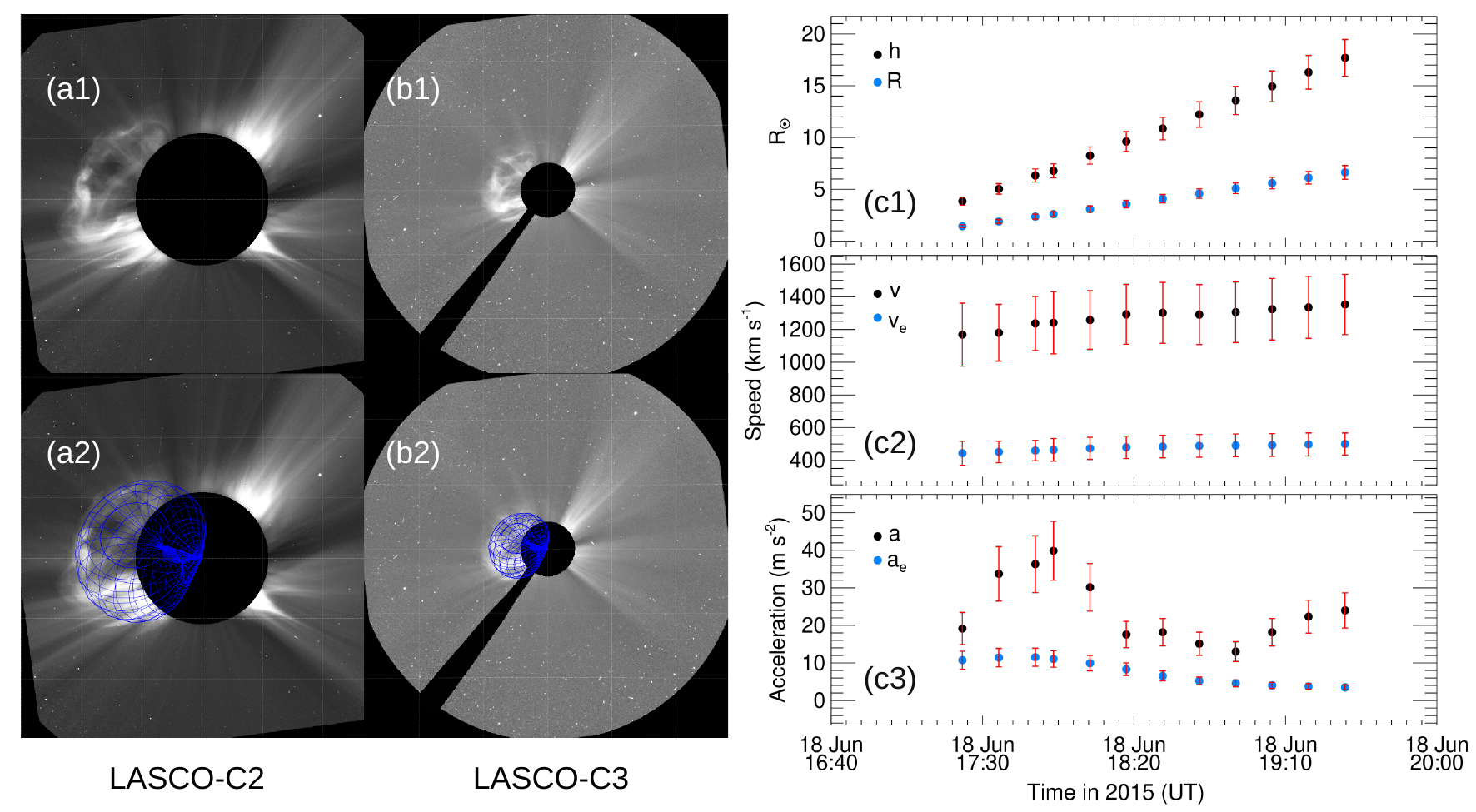}
   \caption{The white light images obtained from the SOHO/LASCO coronagraphs on 2015-06-18, capturing the first CME of CR2165. Subplots (a1) and (b1) show the LASCO C2 and C3 images, respectively. The traced GCS structure is superimposed on these images, indicated by the blue overlay in subplots (a2) and (b2). The temporal evolution of the traced CME's propagation distance (h), speed (v), and acceleration (a) can be observed in subplots (c1), (c2), and (c3), respectively. Corresponding expansion properties (R, v$_e$, a$_e$) are also shown in the right subplots.}
   \label{fig:GCS_fitting}
\end{figure*}

\begin{figure*}
   \centering
   \includegraphics[width = \textwidth]{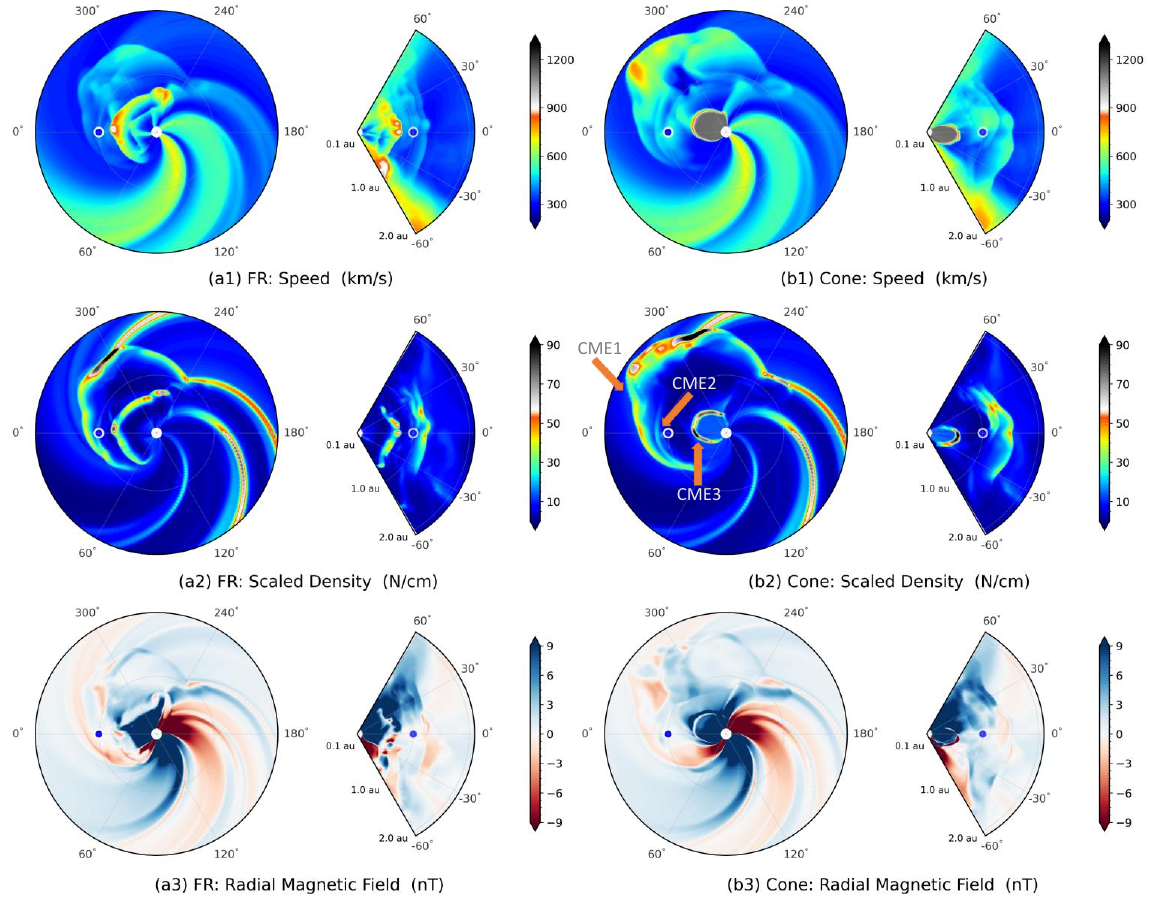}
   \caption{Figure displays a time snapshot of the SWASTi-CME results for CR2165 on 2015-06-22 at 10:26 UT. Subplots (a1)-(a3) depict the speed, scaled density, and radial magnetic field results for the flux rope (FR) CMEs, while subplots (b1)-(b3) show the corresponding results for the cone CMEs. At this particular time, three CMEs, namely CME1, CME2, and CME3, are present in the simulation domain.}
   \label{fig:fr_run}
\end{figure*}

In order to evaluate the performance of SWASTi-CME, we conducted simulations of CME events for two Carrington rotations (CRs) and compared the results with in-situ measurements obtained from the OMNI dataset. The objective was to assess the effectiveness of the model in two different scenarios: one CR with multiple CMEs near solar maxima and another CR with a single isolated CME near solar minima. To achieve this, we specifically chose CR2165, which consisted of five halo CMEs, and CR2238, which featured one halo CME.

\subsection{3D Reconstruction of CME}
To simulate a CME, the foremost step is to reconstruct its 3D geometry in the solar corona using white light images captured by coronagraphs. In this study, we utilized the Graduated Cylindrical Shell \citep[GCS; ][]{thernisien_2011_implementation} forward model for this purpose. Implementing the GCS model requires the manual fitting of a GCS geometry to the white light observations of CMEs to obtain the best model fit parameters. By fitting the model to the temporal sequence of images of CMEs, we could trace the 3D trajectory of the selected CMEs in the corona. This allow us to determine the morphological and kinematic properties of CMEs, such as its de-projected shape, size, and speed. In essence, we employ the GCS technique to obtain the initial properties of the CME at the inner boundary of the MHD domain, specifically at a distance of 21.5 R$_\odot$. These properties include the initial angular width, angular height, tilt, and speed at 21.5 R$_\odot$.  

Figure \ref{fig:GCS_fitting} (a1-b2) presents a snapshot of the GCS fitting applied to white light images obtained from LASCO-C2 and LASCO-C3 coronagraphs. Specifically, the figure showcases the results for the first CME observed during CR2165 (CME1). The blue structures in subplots \ref{fig:GCS_fitting} (a2) and (b2) represent the GCS model fitted wireframe overlaid on the CME. The estimated kinetic properties of CME1 can be observed in subplots \ref{fig:GCS_fitting} (c1), (c2), and (c3), where the three-phase acceleration of the CME \citep{zhang_2004_a}, is clearly visible. The acceleration of the CME initially increases until ~5 R$_\odot$, followed by a decrease and subsequent gradual increase. The same approach is adopted for the other selected CMEs also. Generally, it is recommended to utilize simultaneous coronagraphic observations from multiple viewpoints for confident results from fitting with the GCS model. However, in this case (CR2165), STEREO/COR observations are not available and therefore we could only use SOHO/LASCO observations. However, for CR2238, both STEREO/COR and SOHO/LASCO observations are available, allowing us to utilize contemporaneous images from multiple viewpoints.

In addition to the properties obtained through the GCS method, the estimation of the magnetic properties, density, and temperature of the CME are required. For simplicity, we have assumed a uniform density and temperature for all CMEs and homogeneous speed for cone CMEs. The temperature for each CME was set at 0.8 MK, while the density was estimated through the optimization of the CME's time of arrival at L1. The estimated density was on the order of $10^{-18}$ kgm$^{-3}$, which is consistent with the observations \cite{temmer_2021_deriving}. The magnetic properties of the FRi3D CME were modeled using equation \ref{eq:fri3d_magnetic_equation} with a constant twist, and a magnetic flux of 10$^{12}$ Wb was assumed, which is consistent with values used in \cite{isavnin_2016_fried, singh_2022_ensemble, maharana_2022_implementation}. For the remaining geometric properties of the FRi3D CME, default values were employed. A comprehensive list of all the estimated properties of the CMEs at 21.5 R$_\odot$ is provided in Table 1.

\begin{table*}
\centering
\caption{Initial properties of CMEs associated with CR2165 and CR2238}
\label{tab:cme_properties}
\begin{tabular}{llclclclclclclclc}
\hline
\multicolumn{17}{c}{\textbf{Parameters with different value for each CME}} \\ \hline
\multicolumn{2}{l}{\textit{}} & \multicolumn{1}{l}{Time of arrival at 21.5 R$_\odot$} &  & \textit{v$_{\rm CME}$} &  & \textit{$\theta_{\rm CME}$} &  & \textit{$\phi_{\rm CME}$} &  & \textit{$\varphi_{\rm hw}$} &  & \textit{$\varphi_{\rm hh}$} &  & \textit{$\gamma$} &  & $\rho_{\rm CME}$ \\ \hline
\multicolumn{2}{l}{CME1} & 20:03 UT 18 June 2015 &  & 1353 &  & 6.06 &  & -51.18 &  & 65.3 &  & 36.9 &  & 66.6 &  & 1e-18 \\
\multicolumn{2}{l}{CME2} & 11:25 UT 19 June 2015 &  & 836 &  & -22.5 &  & 7.06 &  & 40.7 &  & 18.7 &  & 32 &  & 2e-18 \\
\multicolumn{2}{l}{CME3} & 05:09 UT 21 June 2015 &  & 1115 &  & 11.03 &  & -24.41 &  & 84.8 &  & 46.8 &  & -85 &  & 2e-18 \\
\multicolumn{2}{l}{CME4} & 21:26 UT 22 June 2015 &  & 1100 &  & 25.11 &  & 7.8 &  & 55.2 &  & 36.1 &  & 68.4 &  & 1e-18 \\
\multicolumn{2}{l}{CME5} & 10:41 UT 25 June 2015 &  & 1330 &  & 23.37 &  & 31.76 &  & 34 &  & 16.2 &  & 77 &  & 3e-17 \\
\multicolumn{2}{l}{CME6} & 19:30 UT 07 Dec. 2020 &  & 1250 &  & -24 &  & 10 &  & 79 &  & 44 &  & 10 &  & 2e-18 \\
 &  & \multicolumn{1}{l}{} &  & \multicolumn{1}{l}{} &  & \multicolumn{1}{l}{} &  & \multicolumn{1}{l}{} &  & \multicolumn{1}{l}{} &  & \multicolumn{1}{l}{} &  & \multicolumn{1}{l}{} &  & \multicolumn{1}{l}{} \\ \hline
\multicolumn{17}{c}{\textbf{Parameters with common value for all CMEs}} \\ \hline
\multicolumn{2}{c}{$T_{\rm CME}$         =    0.8 MK} & \multicolumn{1}{l}{} &  & \multicolumn{3}{c}{$\tau$      =      2} &  & \multicolumn{2}{l}{} &  &  &  & \multicolumn{4}{l}{$\varphi_{\rm p}$=     0.6} \\
\multicolumn{2}{c}{$\phi_{\rm B}$ = 1e12 Wb} & \multicolumn{1}{l}{} &  & \multicolumn{3}{c}{$n$          =     0.5} & \multicolumn{1}{c}{} & \multicolumn{2}{c}{} &  &  &  & \multicolumn{4}{l}{$\varphi_{\rm s}$=     0.0} \\ \hline
\end{tabular} \\
\vspace{1em}
\small Note: ($\theta_{\rm CME}$, $\phi_{\rm CME}$) - CME insertion coordinates; $\varphi_{\rm hw}$ - angular half-width (degree); $\varphi_{\rm hh}$ - angular half-height (degree); $\gamma$ - tilt angle (degree); $\rho_{\rm CME}$ - mass density (kgm$^{-3}$); $\phi_{\rm B}$ - magnetic flux (Wb); $\tau$ - twist; $T_{\rm CME}$ - temperature (MK); $n$ - flattening; $\varphi_{\rm p}$ - pancaking; $\varphi_{\rm s}$ - skew.
\end{table*}

\subsection{MHD Simulation results}
Based on the GCS model results, the first CME of CR2165 (CME1) reached the radial distance of 0.1 AU at 20:03 UT on June 18, 2023, and entered the simulation domain. CME1 was characterized as a fast CME with a speed of 1353 km/s. It had a half angular width of 65.3\textdegree{} and a half angular height of 36.9\textdegree{}. As CME1 started propagating through the interplanetary medium, it encountered a stream interaction region (SIR) along its trajectory. This interaction led to the distortion of the arc structure of the CME front, which is evident from the scaled density plot displayed in Figure \ref{fig:fr_run}. The interaction with the SIR resulted in the formation of a region of elevated density in front of the eastward flank of the CME. Notably, despite both the cone and FR CME initiating simultaneously and propagating through the same medium, they exhibited clear differences in their evolutionary behavior. Specifically, the cone CME exhibited a faster propagation speed compared to the FR CME. This disparity can be attributed to the inherent characteristics of the cone CME, which possessed a homogeneous speed at the onset, while the velocity of the FR CME decreased across its cross-section. Additionally, the cone CME appears to cover a larger area in the $r-\phi$ plane, as shown in Figure \ref{fig:fr_run}. This difference may be caused by the tilt angle of the FR CME, which is not considered in the cone model. Given that the tilt angle of CME1 is 66.6\textdegree{}, its projection onto the equatorial plane will be smaller compared to a CME with no tilt.

Approximately 18 hours after the insertion of CME1, CME2 started entering the simulation domain. CME2, characterized by weaker attributes including a speed of 836 km/s and a notably smaller size compared to CME1, demonstrated an intriguing phenomenon. As CME1 propagated through the inner heliosphere, it displaced the surrounding solar wind plasma, creating a less dense interplanetary medium in its path. Due to which, CME2 experienced lesser drag and was able to intercept the westward flank of CME1 nearly at 1.0 AU. However, the relative speed between the two CMEs was not high, resulting in a relatively weak interaction. In Figure \ref{fig:fr_run}, there aren't very clear signatures of their interaction in $r-\phi$ plane but it can be seen in $r-\theta$ plane, where CME2 has started interacting with CME1. Owing to the less dense medium, CME3 was able to interact with the joint region of CME1 and CME2. This interaction was not as weak as earlier, because CME3 was the largest CME among the five with initial speed of 1115 km/s. This makes it an interesting case, however none of the interaction happened at L1, hence no in-situ signature were recorded in the observations at L1.

CME4 followed a trajectory aligned with the path where the last three CMEs had previously interacted. Despite its initial speed of 1330 km/s, CME4 was unable to catch up with CME3 until reaching a radial distance of 2.0 AU. However, the decreasing distance between CME4 and CME3 indicated an eventual interaction beyond 2.0 AU. In a similar manner, CME5, with a velocity of 1250 km/s, propagated towards the direction of the preceding CME, experiencing less drag. Remarkably, all five CMEs impacted the Earth, with CME3 displaying the highest strength as determined by in-situ measurements. It is worth noting that the successive ejections of CMEs in close proximity led to significant erosion of the SIR, which was located along the trajectory of CME1. This SIR could only regained its heliospheric structure after the transit of CME5.

The CME of CR2238 (CME6) was also a fast one, with initial speed of 1250 km/s, and and significant size, with a half angular width of 79\textdegree{} and a half angular height of 44\textdegree{}. It was the only halo CME observed during this Carrington rotation period, and its trajectory did not encounter any prominent SIR. Unlike the CMEs of CR2165, CME6 did not have a direct impact on Earth; instead, it merely grazed as it propagated through the inner heliosphere.

\subsection{Validation with in-situ measurements}

\renewcommand{\thesubfigure}{}
\begin{figure*}
  \centering
  \subfigure[]{\includegraphics[width = \textwidth]{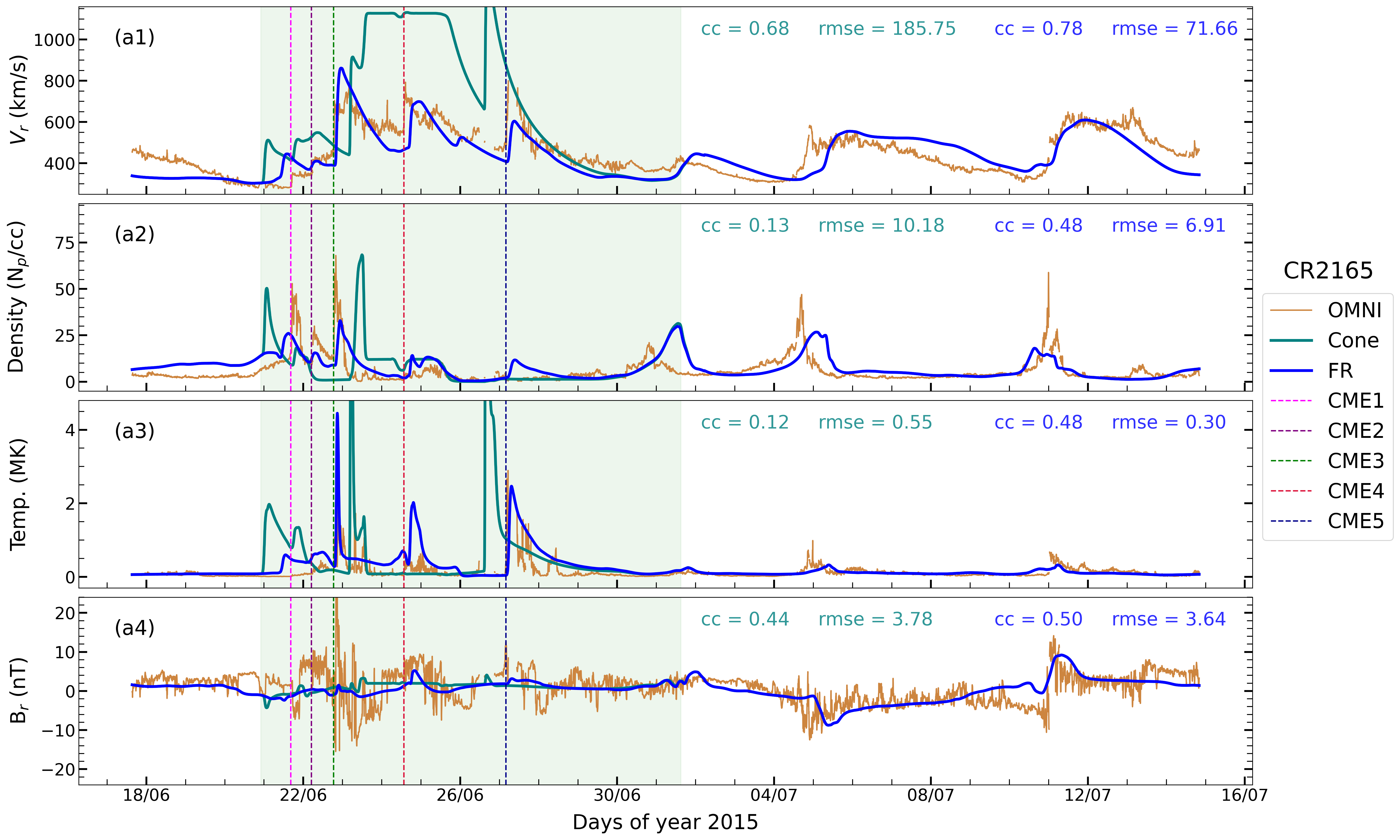}}\vspace{-2em}
  \subfigure[]{\includegraphics[width = \textwidth]{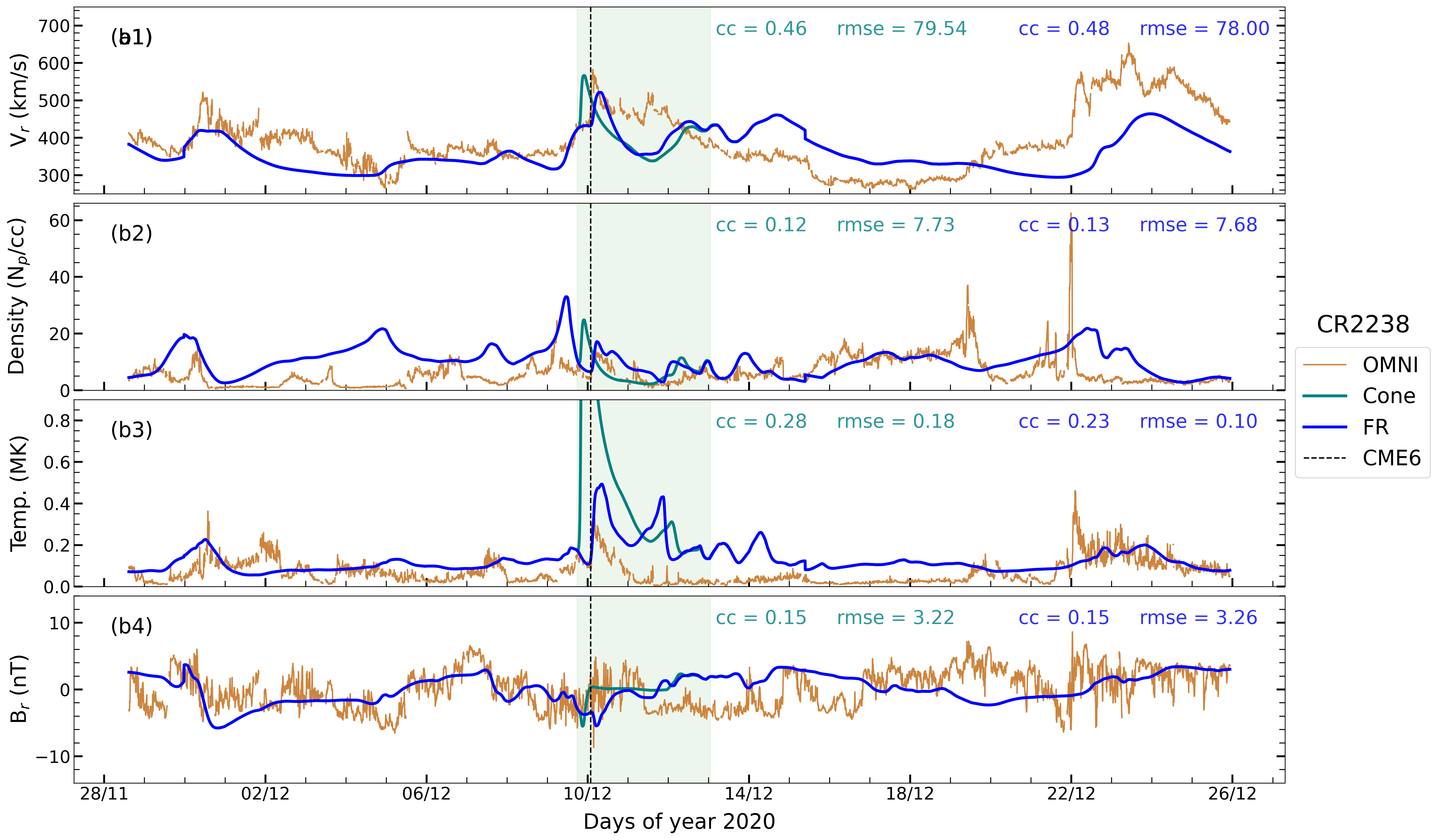}}\vspace{-1em}
    \caption{Time series plot of in-situ measurements at 1 AU for CR2165 (a1-a4) and CR2238 (b1-b4). The simulated results for the FR and cone CMEs are represented by blue and green colors, respectively, while the observed OMNI data is depicted in brown. Vertical dashed lines of various colors indicate the beginning of the sheath region corresponding to each CME. Both the SWASTi and OMNI data are plotted with a time cadence of 5 minutes.}
    \label{fig:CR2165}
\end{figure*}                       

To evaluate the precision of the SWASTi-CME model, we conducted a comparative analysis between the simulation results and the OMNI data of CR2165 and CR2238. To ensure a fair comparison with the L1 measurements, a virtual spacecraft was integrated into the simulation, which effectively considered the dynamic changes in the longitudinal and latitudinal position of Earth in the Stonyhurst coordinate system. This virtual spacecraft facilitated the acquisition of plasma properties at 5-minute intervals throughout the computational run, enabling a direct comparison with the 5-minute average data obtained from OMNIWeb.

Figure \ref{fig:CR2165} presents the time series plots illustrating the cone and FR CMEs, along with the OMNI data for CR2165 and CR2238. The FR CME results are represented by the blue solid line, the cone CME results are shown in green, and the OMNI data is depicted in brown. The shaded regions indicate the temporal range where the cone and FR CME results differ, highlighting the distinction between the ambient SW and the CME region. The vertical dashed lines in various colors mark the time of arrival of the sheath region for different CMEs.

\textit{Time of Arrival.} In the case of CR2165, the FR CMEs successfully predicted the arrival time of all five CMEs with good accuracy. However, the cone CMEs, although reaching the Earth's location, exhibited less precise estimations of the time of arrival. The velocity (a1) and density (a2) subplots in Figure \ref{fig:CR2165} demonstrate this discrepancy. Cone CME1 arrived much earlier than the observed time (-15.43 hr), along with CME2 (-9.75 hr) and CME3 (-15.84 hr), compared to FR CME1 (-3.85 hr), CME2 (+1.42 hr), and CME5 (+3.67 hr). Conversely, cone CME3 arrived later (+19.4 hr) than FR CME3 (+6.0 hr), while cone CME4 estimated a better time of arrival (+3.83 hr) than FR CME4 (+4.92 hr). For CR2238, both FR and cone CMEs provided efficient estimations of the time of arrival, with the cone CME arriving earlier than the observed time and the FR CME arriving later.

\textit{Speed.} Similar to the time of arrival, the computed CME speed was more accurate for FR CMEs compared to cone CMEs in the case of CR2165. The FR CMEs exhibited a Pearson correlation coefficient (cc) of 0.78 with observations of full CR2165 period and a root mean square error (rmse) of 71.66 km/s. On the other hand, the cone CMEs had a cc of 0.68 and a rmse of 185.75 km/s when compared to the observations. The velocity results from the FR model closely aligned with the observed values for the first four CMEs. However, the simulated speed of CME5 was underestimated by approximately 200 km/s at 1 AU. On the other hand, the cone model overestimated the speed for all CMEs and particularly provided very high values for CME3, CME4, and CME5 compared to the observed values. This discrepancy can be attributed to three factors: homogeneous speed, insertion rate, and the absence of tilt in the cone model. The cone CME, with a constant speed throughout its structure, had a higher effective momentum than the flux rope CME. Moreover, the injection rate of the cone CME, proportional to $\tan(\varphi_{\rm hw})$, was slower for CMEs with higher angular width ($>$75\textdegree{}) compared to a CME with an angular width of 60\textdegree{}. This delay in insertion ultimately impacted the time of arrival at 1 AU. Hence, even with a higher homogeneous speed, CME3 arrived later due to its significantly high angular half-width (84.8\textdegree{}). Additionally, since the cone CME model does not incorporate tilt, the effective trajectory traced by the CME before reaching 1 AU differs from that of the FR CME, which has a higher degree of tilt. For CR2238, both the cone (cc=0.46, rmse=79.54 km/s) and FR (cc=0.48, rmse=78.00 km/s) models provided good estimations of the CME speed at 1 AU. In this case, where CME didn't directly impact the Earth but instead grazed by, the speed estimation in the cone CME was also appropriate.

\textit{Density and Temperature.} In contrast to FR CMEs, the cone CMEs exhibited higher density and temperature in their sheath regions for all six CMEs. This can be attributed to the homogeneous speed and absence of tilt angle in the cone CME model, which leads to a larger surface area and higher speed in the elliptic plane, consequently resulting in hotter and denser sheath regions as the cone CME propagates in the heliosphere. This pattern is evident in subplots (a2) and (a3) of Figure \ref{fig:CR2165}. Moreover, CME4 and CME5 presented an interesting scenario where the CME speeds were relatively higher, but the density values were lower. This can be explained by the influence of CME3, which had a very high angular width. The propagation of CME3 likely depleted the density of the interplanetary medium, limiting the aggregation of plasma and resulting in less dense sheath regions for CME4 and CME5.

\textit{Magnetic Field.}  There is noticeable difference in the in-situ magnetic profile between cone and flux rope CMEs. In Figure \ref{fig:fr_run} (a4) and (b4), the radial magnetic field remains almost constant with time for cone CMEs, while for flux rope CMEs, it closely matches the feature observed at L1. This indicates that the flux rope CMEs exhibit better agreement with the observed data. However, it is worth noting that the estimated magnitude of the magnetic fields did not always match perfectly with the in-situ data. This suggests that utilizing a constant magnetic flux value for each CME is not the most suitable choice. Adopting different values for different CMEs might lead to improved accuracy in predicting magnetic field variations for the considered cases.

\textit{Solar Wind.} Regarding the properties of the ambient solar wind, the SWASTi model successfully captured all major features for both CR2165 and CR2238. The simulated speeds, densities, temperatures, and magnetic fields of the solar wind closely aligned with the measurements obtained from the OMNI dataset. This demonstrates the efficacy of the solar wind model in generating a near-realistic ambient environment for studying the interactions with CMEs even at a high cadence of 5 minutes.

\section{Method to study CME-SW Interaction} \label{section4}
As a CME propagates through the inner heliosphere, it may undergo changes in its dynamics and morphology due to its interaction with the ambient solar wind (SW). Variations in the properties of the ambient SW, such as its speed and density, can result in alterations to the properties of the CME and its shock \citep{winslow_2021_the, mishra_2021_radial}. Investigating the interaction between coronal mass ejections (CMEs) and solar wind (SW), especially in the presence of high-speed streams (HSS) and stream/corotating interaction regions (SIRs/CIRs), can be accomplished by comparing two scenarios in which an identical CME passes in the presence and absence of them. However, to comprehensively evaluate the impact of SW on CME, the entire CME structure must be isolated to measure the differences between the scenarios accurately. In this numerical study, we utilized the SWASTi framework to simulate two different ambient solar wind scenarios. We introduced a flux rope CME into both mediums and employed a novel technique for tracking the 3D CME structure as it propagated through the heliosphere, enabling us to quantify any changes that occurred.

In this section, we discuss the numerical setup that was used to study the effect of SW on CMEs. We implemented this setup on two different Carrington rotation periods, namely CR2165 and CR2238, which were previously validated in Section \ref{section3}. 

\subsection{Numerical setup}

\begin{figure*}
   \centering
   \includegraphics[width = \textwidth]{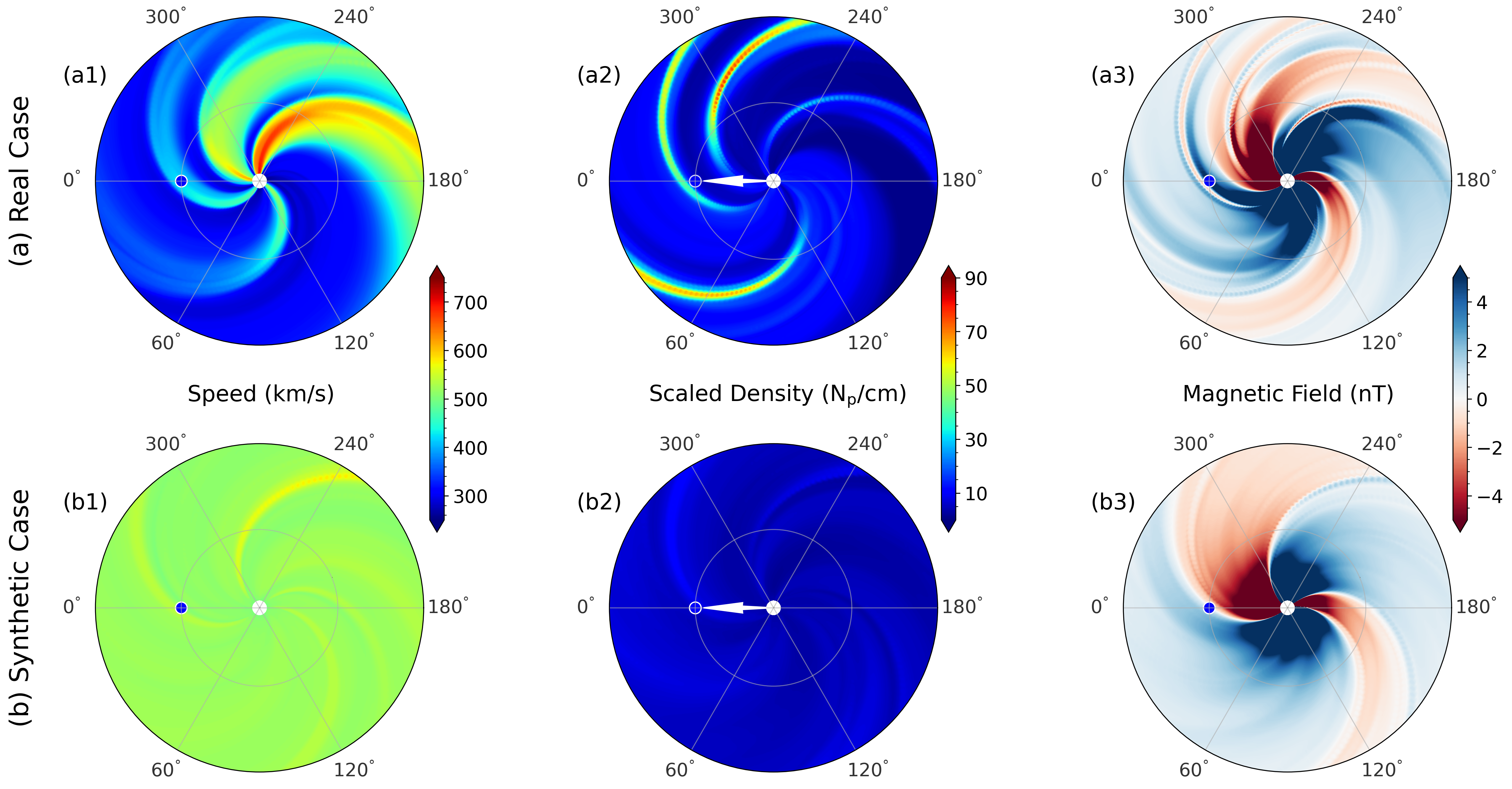}
   \caption{The figure showcases the equatorial plane of the numerical setup used to study the influence of ambient solar wind on a coronal mass ejection (CME). Subplots (a1)-(a3) present the speed, scaled density, and radial magnetic field for the real case of CR2165, while subplots (b1)-(b3) depict the results of the synthetic case.}
   \label{fig:real_vs_ideal}
\end{figure*}

To examine the changes in CME properties caused by the ambient SW, we established two setups. The first setup (\textit{real} case) is the default setup of the SWASTi-CME flux rope model and has the realistic SW background. Whereas, the second setup (\textit{synthetic} case) is a hypothetical case where solar wind speed is assumed to be almost constant everywhere in the heliosphere. In other words, all the features formed due to the relative speed between different SW streams, for e.g., HSSs and SIRs/CIRs, will be present in the real case but absent in the synthetic case. Further, an identical CME is allowed to propagate in both the cases. Passing through a realistic SW environment, the real case CME will encounter all the variations caused by HSSs and SIRs/CIRs that the synthetic case CME will not. Therefore, the difference in the properties of CMEs corresponding to real and synthetic cases, will directly reflect the impact of HSSs and SIRs/CIRs.

This two setup methodology was carried out for CR2165 and CR2238. The real case SW background simulation, for both CRs, is identical to what is described in Section 3. In the synthetic case, the initial boundary conditions at 0.1 AU are different, where a constant value of the solar wind speed is provided. In this study, this value is taken to be close to the mean of real case SW speed on the equatorial plane, ensuring similar propagation duration for both real and synthetic CMEs in the simulation domain. At the initial boundary, the density and pressure profiles are also different since they are both dependent on speed, but the magnetic field is kept identical for both the cases. Table \ref{tab:tabel1} shows the values of the constant speed taken for the synthetic case of CR2165 and CR2238, which is 500 km/s and 450 km/s respectively. 

A flux rope CME structure, with the properties as mentioned in Table \ref{tab:tabel1}, is injected in both the cases of CR2165 and CR2238. The intention was to deploy a CME with simple structure and symmetry about the equatorial plane, so that the effects can be visualized easily. Therefore, 0$^\circ$ latitude and 0$^\circ$ longitude was taken as the center of CME, with mass density = 10$^{-18}$ kgm$^{-3}$, front speed = 900 kms$^{-1}$, temperature = 0.8 MK, and magnetic flux = 10$^{12}$ Wb. Additionally, the half-width, half-height and tilt angle was 45$^\circ$, 20$^\circ$ and 0$^\circ$, respectively. Although the above CME features were same for real and synthetic cases of both CRs, the injection time of the CMEs were different. Since, there is significant difference in the SW properties of CR2165 and CR2238, it is important to choose a sensible CME injection time favorable for a significant interaction between CME and ambient SW. The insertion time will determine the local SW surrounding and effective trajectory of CME in the heliosphere, which in turn, will determine the degree of deformation experienced by the CME under the influence of pre-conditioning of the ambient SW structures. CR2165 has a greater variation in SW speed and stronger SIRs than CR2238. Therefore, to encompass a wider range of impacts on the CME, an appropriate injection time was selected to ensure that the CME propagated towards the SIR. Specifically, for CR2165, the direction was towards two high density (strong) SIRs, and for CR2238, it was towards a comparatively lower density (weaker) SIR. The injection location can be seen in (a2) subplot of Figure \ref{fig:real_vs_ideal} (marked with white arrow) for CR2165 which corresponds to 10:23 UT 02 July 2015. And similarly, 12:16 UT 20 November 2020 was the CME injection time for CR2238.

\begin{table}
\centering
\caption{Simulation parameters for the synthetic and real cases
of CR2165 and CR2238.}
\label{tab:tabel1}
\begin{tabular}{lllllllllll}
\hline
\multicolumn{11}{c}{CME parameters for Real and synthetic Cases} \\ \hline
$\theta_{\rm CME}$ & = & 0$^\circ$ &  &  &  &  &  & $n$ & = & 0.5 \\
$\phi_{\rm CME}$ & = & 0$^\circ$ &  &  &  &  &  & $\varphi_{\rm p}$ & = & 0.6 \\
$\varphi_{\rm hw}$ & = & 45$^\circ$ &  &  &  &  &  & $\varphi_{\rm s}$ & = & 0 \\
$\varphi_{\rm hh}$ & = & 20$^\circ$ &  &  &  &  &  & $\phi_{\rm B}$ & = & 10$^{12}$ Wb \\
$\gamma$ & = & 0$^\circ$ &  &  &  &  &  & $\tau$ & = & 2 \\
$\rho_{\rm CME}$ & = & 10$^{-18}$ kg m$^{-3}$ &  &  &  &  &  & $T_{\rm CME}$ & = & 0.8 MK \\
 &  &  &  &  &  &  &  &  &  &  \\ \hline
\multicolumn{11}{c}{Solar Wind properties for synthetic case} \\ \hline
CR2165 &  &  &  &  &  &  &  & \multicolumn{3}{l}{CR2238} \\
$v_{\rm in}$ & = & 500 &  &  &  &  &  & $v_{\rm in}$ & = & 450 \\
$P_{\rm in}$ & = & 6.0 nPa &  &  &  &  &  & $P_{\rm in}$ & = & 6.0 nPa \\
$\rho_{\rm fsw}$ & = & 200 &  &  &  &  &  & $\rho_{\rm fsw}$ & = & 200 \\ \hline
 &  &  &  &  &  &  &  &  &  & 
\end{tabular}
{\raggedright {\small Note: ($\theta_{\rm CME}$, $\phi_{\rm CME}$) - CME center coordinates; $\varphi_{\rm hw}$ - angular half-width; $\varphi_{\rm hh}$ - angular half-height; $\gamma$ - tilt angle; $\rho_{\rm CME}$ - mass density; $n$ - flattening; $\varphi_{\rm p}$ - pancaking; $\varphi_{\rm s}$ - skew; $\phi_{\rm B}$ - magnetic flux; $\tau$ - twist; $T_{\rm CME}$ - temperature; $v_{\rm in}$ and $P_{\rm in}$ - SW speed and pressure at 0.1 AU; $\rho_{\rm fsw}$ - SW number density corresponding to 650 km s$^{-1}$ SW stream.} \par}
\end{table}

Figure \ref{fig:real_vs_ideal} shows the SW simulation results of real and synthetic cases for CR2165. The figure demonstrates the fundamental differences between the real case (subplots a1, a2 and a3) and synthetic case (subplots b1, b2 and b3) in the equatorial plane. In the synthetic case, the SW speed is almost constant everywhere and hence there's no formation of HSS and SIR/CIR structures. Just like the speed, the scaled density is also relatively uniform in longitudinal direction as compared to the real case. And the magnetic field subplot displays a simple magnetic dipole configuration, whereas for the real case it is more complex due to interaction among solar wind streams of differing characteristics. The spiral features visible in synthetic case subplots (b1) and (b2) are corresponding to the polarity inversion line in (b3). Due to very low magnetic field strength in those regions, the SW density is lower and hence, speed is slightly higher.

\subsection{CME structure identification} \label{section:CME_isolation}

\begin{figure*}
   \centering
   \includegraphics[width = \textwidth]{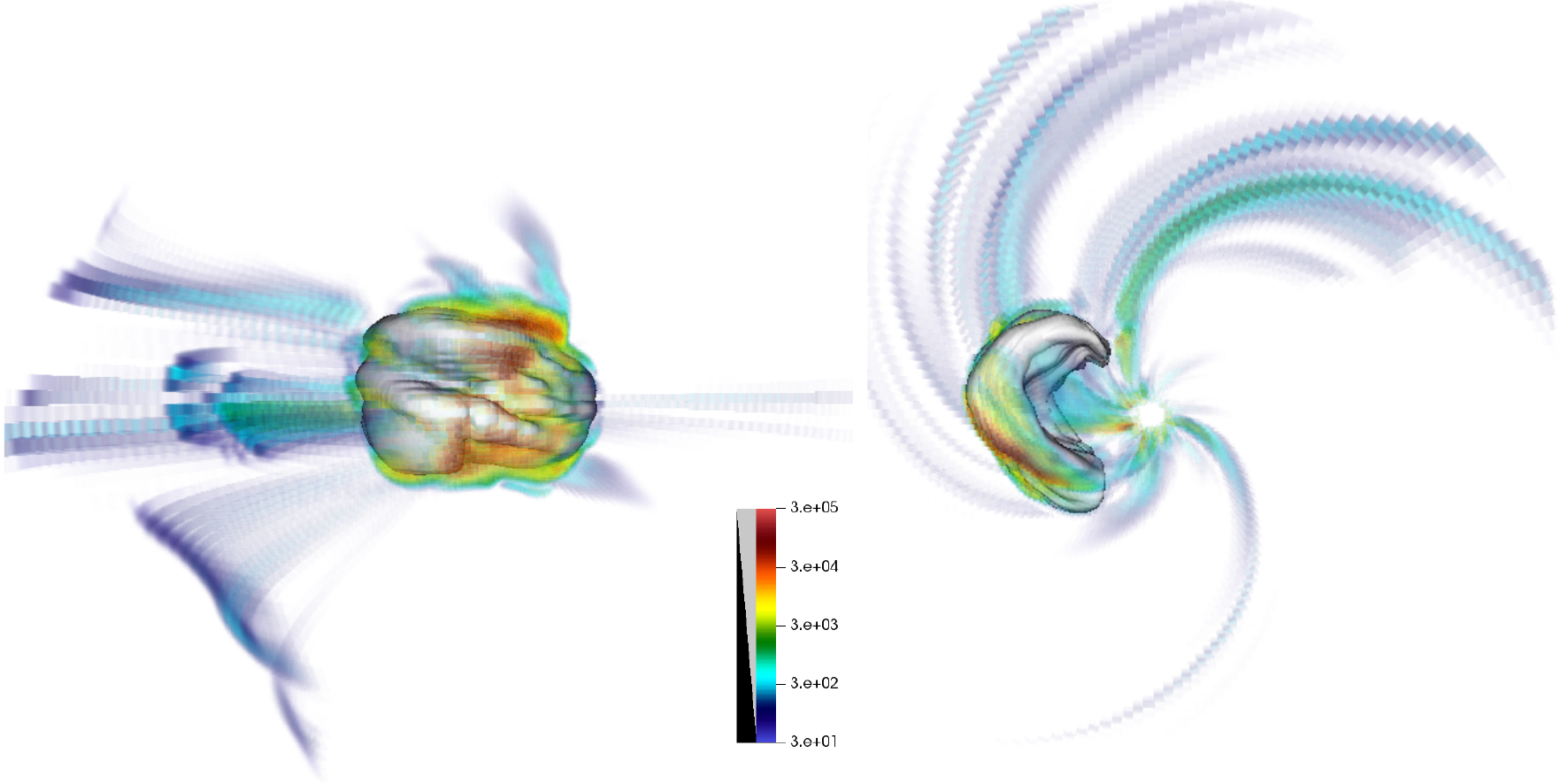}
   \caption{Figure demonstrates the tracing of the three-dimensional structure of a CME (denoted by white) in the heliosphere. This volumetric plot showcases the changes in dynamic pressure of the solar wind (denoted by colorbar) caused by the propagation of the CME. The left subplot displays the deformation of the CME front resulting from the interaction with a SIR.}
   \label{fig:3D_2165}
\end{figure*}

It is expected that structure of CME can be deformed during its propagation in the structured solar wind. Exploration of the heliospheric imagers observations have provided observational evidence that such deformation in CME is significant during its interaction with with SIRs and other preceding CMEs \citep{liu_2012_interactions, mishra_2014_a, heinemann_2019_cmehss}. The preceding structures in the background solar wind play a role of density/field obstacles and also cause a steep gradient in the solar wind speed for the following CME. The effect can be more pronounced at large distances from the Sun where the momentum of interacting structures dominates the magnetic force. To identify and track the evolution of whole CME structure in our simulations, we have adopted a method of tracer. This CME tracer ($\mathcal{T}_r$) is a standard passive scalar that follows a simple advection equation and does not engage in any kind of interaction with any other physical quantities. It is initialized at the time of CME insertion with the value unity inside the CME structure and null elsewhere in the computational domain. This passive scalar can be visualized as a `color' associated with the CME that follows with its bulk motion and as the CME mixes with the ambient solar wind, the value of this `color' fades. In this work, the $\mathcal{T}_r$ lower limit to define the CME structure is set to 0.1 and following is the conservative form of the equation that governs it:

\begin{equation} \label{eq:tracer}
    \frac{\partial (\rho \mathcal{T}_r)}{\partial t} \quad + \quad \nabla \cdot (\textbf{v} \rho \mathcal{T}_r)\quad = \quad 0
\end{equation}

Similar CME tracking method was also used by \cite{biondo_2021_tracing} to show that CME bubble remains intact even after interaction with multiple solar wind streams.

Figure \ref{fig:3D_2165} demonstrates the identification of the CME structure using the aforementioned scheme. The CME structure, represented in white color, is traced in the heliosphere, while the volumetric plot displays the ambient solar wind streams, with the color map indicating the dynamic pressure of the solar wind. To enhance visibility, the upper and lower limits of dynamic pressure ($\rho_{sw}v_{sw}^2/2$) have been adjusted to highlight relatively strong SIRs. The left subplot showcases the front view of CME propagation, while the right subplot presents a top view. It is worth emphasizing that the traced CME's front exhibits some deformities rather than being smooth. A better understanding of deviations in the structural front of a CME from its assumed rigid and designated shape due to gradient in the ambient medium properties can help improve its arrival time on Earth. Further analysis of such characteristics is discussed in detail in the subsequent sections.

\section{CME-Solar Wind Interaction} \label{section5}

In this section we present the results of the simulations described in section \ref{section4}, discussing the interaction of CME with structured solar wind (SW) and its impact on the evolution of CME. We focus on investigating the changes in the morphological, dynamic, and physical characteristics of CMEs as they propagate through a non-uniform ambient SW and encounter SIRs.

\subsection{Evolution of CME structure} \label{section3.1}

\begin{figure*}
   \centering
   \includegraphics[width = \textwidth]{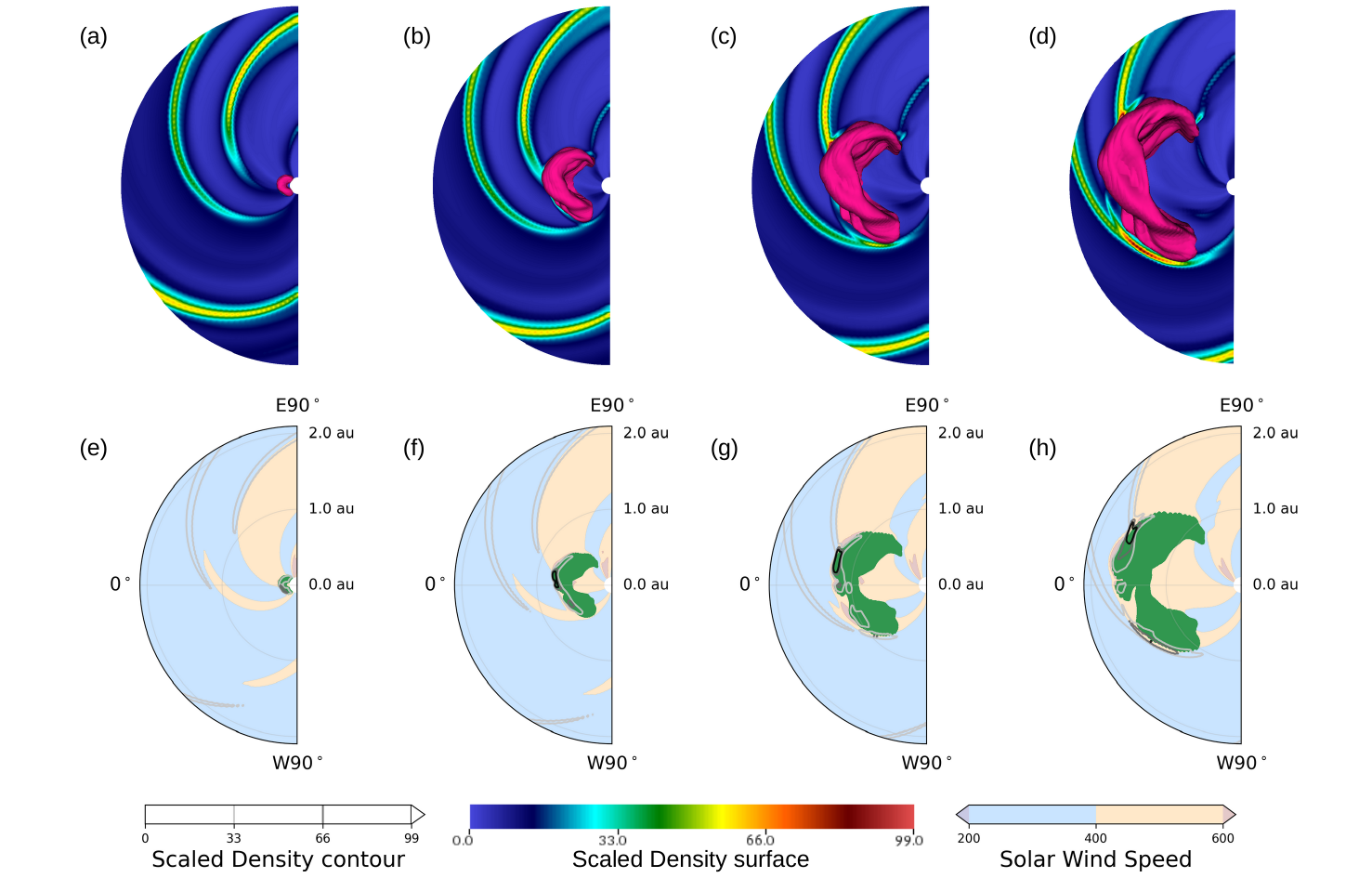}
   \caption{The evolution of real case CME in the heliosphere and its interaction with ambient solar wind of CR2165 is depicted in four time steps. Top row (a)-(d): represent the ecliptic plane, showcasing surface plot of the scaled density of the solar wind with the 3D CME (depicted in pink) propagating through it. Bottom row (e)-(f): display a 2D slice of the 3D CME structure (shown in green), along with the surface plot of the solar wind speed, and line contours represent the scaled density.}
   \label{fig:4-step_evolution}
\end{figure*}
    
The interaction of CME with a SIR can result in observable change in the shape of the CME front making it deviate from the self-similar expansion and can also affect the global trajectory of the CME \citep{winslow_2021_the}. An overview of the CME propagation in the presence of HSS and SIR (\textit{real case} of CR2165) is demonstrated in Figure \ref{fig:4-step_evolution}, where the top row depicts the 3D evolution of the CME structure (pink) and the bottom row shows the 2D slice of same CME (green) in ecliptic plane. Each column represents different time snapshots of the CME evolution in the inner heliosphere. The subplots (a), (b), (c) and (d) have surface plot of scaled density in the background, whereas subplots (e), (f), (g) and (h) have contour plot of scaled density along with the speed of solar wind.  

\textbf{Deflection.} Our simulation shows that during the initial stage of CME evolution, the movement of the CME in the $r-\phi$ plane is slightly directed towards the eastward direction, as evident from the Figure \ref{fig:4-step_evolution}(f). The spiral structure of SW streams in the heliosphere, which propagates from the west to the east, can obstruct the westward motion of CME. This hindrance is more prominent when the CME encounters high-density SIRs. When the SIR is strong enough, it can impel the CME to follow the trajectory of the SIR, resulting in the CME's eastward movement, as illustrated in subplot (f). The extent of deflection experienced by a CME is expected to depend on  the relative strength of the SIR and the CME. We could notice that for the real case of CR2165, the SIR was not strong enough to contain the CME for an extended period of time, and thus the deflection is primarily noticeable in the initial phase. Moreover, the interaction between the CME's eastward flank and the SIR has resulted in a coarser CME surface compared to the smoother westward flank, which did not come into contact with the SIR. This asymmetry in CME surface suggests that the interaction with the SIR can have a significant impact on the physical characteristics of the CME.

\textbf{Expansion.} In Figure \ref{fig:4-step_evolution}, the CME injected into the simulation domain was subjected to faster solar wind streams on its westward flank and slower streams on the eastward flank. The difference in speed also led to a disparity in pressure gradient and drag experienced by the two flanks of the CME due to the higher density and pressure of the slower streams. Consequently, the expansion rates of the westward and eastward flanks of the CME were significantly different. The westward flank of the CME has been observed to have over-expanded, whereas the eastward flank remained under-expanded due to direct contact with the SIR. This variation in the expansion rates along the CME structure will result in the change of density distribution of CME. In essence, speed variations in the ambient solar wind may introduce substantial changes to the density distribution of the CME.

\textbf{Leading Edge.} The presence of a strong SIR is primarily observed near the equatorial region, as slower or denser solar wind streams are often confined in that region \citep{gosling_1999_formation, richardson_2018_solar}. As the CME starts propagating in the heliosphere, its sheath region typically displays the highest  density, with a nearly symmetric distribution around the CME's axis of propagation. However, as the CME propagates further, it is possible that the flank of the sheath regions that come into contact with the SIR exhibit the highest density values. For instance, in subplots (e) and (f), the density is highest at the front of the CME, while having the CME at larger distance in subplot (h), the density is highest at the sides. Moreover, the regions with the higher density values tend to be slightly pushed behind by the surrounding ambient solar wind, as evident in the eastward flank of subplots (g) and (h). This lagging effect is attributed to the greater effective drag force  experienced by these regions compared to other parts of the CME, causing them to be pushed rearward. Consequently, the CME's leading edge exhibits a non-uniform or non-circular shape due to the varying solar wind conditions and the dynamic drag force encountered along its front.

\subsection{Evolution of CME properties}
In past, the evolution of CME has been investigated through observations \citep{liu_2005_a, janvier_2019_generic, salman_2020_radial} as well as through simulations \citep{manchester_2017_the, scolini_2021_exploring}. Typically, the variation of CME properties is investigated through a power law function of radial distance, with an anticipated decrease in the magnitude of CME properties over a certain range of exponent values. However, the influence of ambient solar wind conditions on the evolutionary behavior of CME properties is not yet well understood.

In order to investigate this impact on the evolution of CME properties, we compared the temporal variation of median values between real and synthetic cases using the function given below:

\begin{equation} \label{eq:log-log}
    \log \tilde{Q}(t) \quad = \quad \alpha_q \cdot \log(t) + \quad \log(Q_0) 
\end{equation}
\begin{equation} \label{eq:slope_vs_time}
    slope = \frac{\log \tilde{Q}_{i+1} - \log \tilde{Q}_{i}}{\log t_{i+1} - \log t_{i}}
\end{equation}

\begin{table}
\centering
\caption{The values and the range of power law followed by different CME parameters.}
\label{tab:power_law}
\begin{tabular}{@{\rule{0pt}{15pt}}lcccc@{}}
\toprule
\multirow{2}{*}{Parameters} & \multicolumn{2}{c}{CR2165} & \multicolumn{2}{c}{CR2238} \\ \cmidrule(l){2-5} 
& Real & Synthetic & Real & Synthetic \\ \midrule
$\alpha_{\rm th}$ & -2.30$^{+0.54}_{-0.37}$ & -3.15$^{+0.40}_{-0.36}$ & -2.50$^{+0.23}_{-0.1}$ & -3.34$^{+0.19}_{-0.34}$ \\
$\alpha_{\rm mag}$ & -2.37$^{+0.51}_{-0.24}$ & -2.76$^{+0.18}_{-0.26}$ & -2.24$^{+0.31}_{-0.28}$ & -2.81$^{+0.23}_{-0.34}$ \\
$\alpha_{\rm ram}$ & -2.09$^{+0.26}_{-0.21}$ & -2.54$^{+0.19}_{-0.15}$ & -2.03$^{+0.26}_{-0.18}$ & -2.53$^{+0.13}_{-0.19}$ \\
$\alpha_{\rm v}$ & -0.1$^{+0.03}_{-0.02}$ & -0.02$^{+0.01}_{-0.01}$ & -0.06$^{+0.01}_{-0.01}$ & -0.02$^{+0.02}_{-0.02}$ \\
$\alpha_{\rho}$ & -1.88$^{+0.21}_{-0.34}$ & -2.59$^{+0.29}_{-0.14}$ & -1.92$^{+0.18}_{-0.33}$ & -2.55$^{+0.12}_{-0.20}$ \\
$\alpha_{\rm T}$ & -0.46$^{+0.36}_{-0.15}$ & -1.48$^{+0.42}_{-0.31}$ & -0.59$^{+0.13}_{-0.09}$ & -1.42$^{+0.25}_{-0.55}$ \\ \bottomrule
\end{tabular}
\end{table}

    \begin{figure*}
       \centering
       \includegraphics[width = \textwidth]{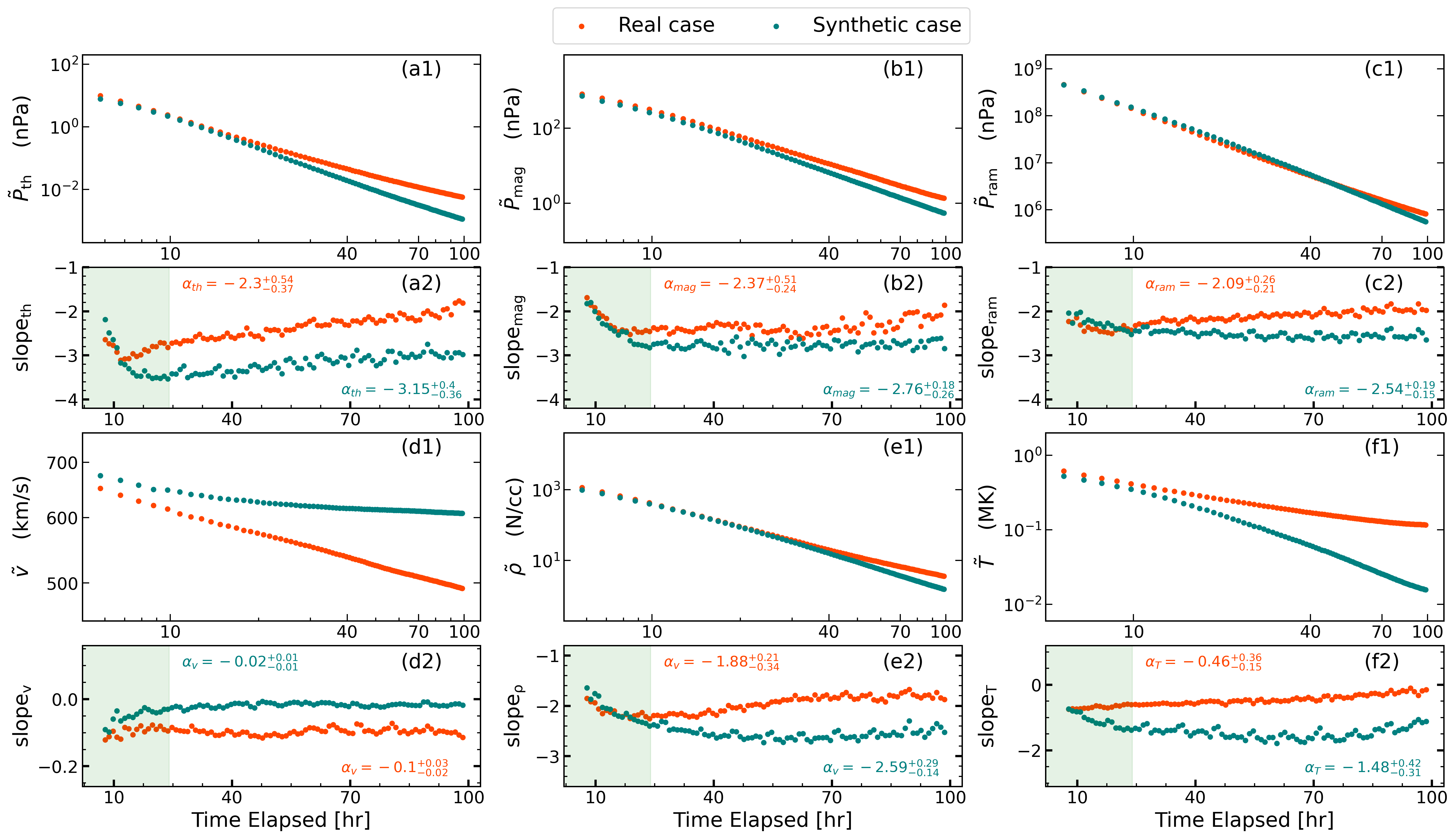}
       \caption{Temporal evolution of median of the CME properties for CR2165. The panels (a1) to (f1) represent the temporal profile of median of CME thermal pressure (a1), magnetic pressure (b1), ram pressure (c1), speed (d1), density (e1), and temperature (f1) in log-log scale. In the subplots (a2) to (f2), just below the panels showing variations in the median of CME properties, the change in the gradient of these CME properties in log-log scale at different time steps are shown (see Equation \ref{eq:slope_vs_time}). The shaded region in slope subplots are corresponding to the initial settling phase of the CME, which has been taken to be 24 hours.}
       \label{fig:median_evolution_2165}
    \end{figure*}

where, $\tilde{Q}$(t) is the median of CME property Q, $t$ is the time, Q$_0$ is the typical value of Q at 1.0 AU, $\alpha_q$ is the power law exponent for $Q$, $i$ is the time step index. As mentioned in last subsection, the location of the CME nose changes continuously as it propagates through the heliosphere, indicating that the rate of change of radial distance is not constant, but rather fluctuating. Additionally, since real and synthetic CMEs have different propagation speeds in the heliosphere, their radial distance profiles will also differ. Therefore, to ensure a fair comparison between the two cases, we opted to fit a power law function of time. While most studies have used mean values with respect to radial distance, for this investigation, examining the time evolution of median values is more appropriate. \cite{scolini_2021_exploring} and \cite{janvier_2019_generic} have also emphasised the suitability of using the median to examine the evolution in their respective studies.

Figure \ref{fig:median_evolution_2165} displays the outcomes of the real and synthetic cases for CR2165. To have a robust comparison, the slope between two points of the evolution plot in a log-log scale was calculated using equation \ref{eq:slope_vs_time}. Subsequently, a straight line parallel to X-axis was fitted to the gradient plot to analyze the power law behavior.  To prevent any bias in the fitting process, we performed it after the initial settling phase of the CME, as including the points immediately after the CME injection could have affected the fitting results. We considered a settlement time of 24 hours for this purpose. Table 2 reports the range of  $\alpha_q$  for both CRs which we got after fitting the slope plots and we found that our results were consistent with the values reported by \cite{liu_2005_a}, \cite{salman_2020_radial} and \cite{scolini_2021_exploring}.

\textbf{Pressures.} The median values of thermal and magnetic pressure decreases with heliospheric distance for both real and synthetic cases of CR2165 and CR2238. For CR2165 (CR2238), the synthetic case CME was found to show more decrement. The fall of magnetic pressure in the synthetic CME followed a power law of $\alpha_{\rm mag}$ = -2.76 (-2.81), while thermal pressure followed $\alpha_{\rm th}$ = -3.15 (-3.34). For real CME, we noticed that the nature of fall of pressures are almost the same as for synthetic case, but with different values: $\alpha_{\rm mag}$ = -2.37 (-2.24) and $\alpha_{\rm th}$ = -2.3 (-2.5). The similarity in the slope variation between the real and synthetic cases, although with different $\alpha$ values, implies that the ambient solar wind condition does indeed significantly impact the evolution of CME pressures. However, this difference mainly manifests during the early stage of CME propagation, and its impact remains consistent over time. Moreover, the larger difference in $\alpha_{\rm th}$ suggests that the interaction has a greater impact on the thermal pressure as compared to the magnetic pressure of CME.

\textbf{Speed.} The median of CME speed in CR2165 (CR2238) slightly decreases with time with a slope of $\alpha_{\rm v}$ = -0.02 (-0.02) for synthetic case and $\alpha_{\rm v}$ = -0.1 (-0.06) for real case CME. Unlike all other CME properties, the speed of CMEs in the both CR2165 and CR2238 exhibits a more prominent decline in the real case as compared to the synthetic case. This can be attributed to the increased amount of drag experienced by the real case CMEs, which is a consequence of their interaction with the non-uniform ambient SW. Additionally, it is worth noting that the fall in speed of CME in CR2165 is more pronounced due to the presence of stronger SIRs in the path of the CME compared to that in CR2238. The increased interaction between the CME and the denser SIRs in CR2165 leads to a more significant deceleration effect, resulting in a steeper fall in speed compared to the CME in CR2238. Additionally, the synthetic CME seems to achieve a terminal speed at the end of the plot, i.e., after 100 hours of propagation in inner-heliosphere.

\textbf{Density.} The CME median proton density falls with $\alpha_{\rho}$ = -2.59 (-2.55) for synthetic case and $\alpha_{\rho}$ = -1.88 (-1.92) for real case of CR2165 (CR2238). In addition to the different $\alpha_{\rho}$ values, the real and synthetic case CMEs also exhibit distinct patterns of slope variation in both CRs. This suggests that the ambient SW conditions have a continuing and accumulating impact on the density distribution as CME propagates through the inner-heliosphere. The over-expansion of one flank of CME could lead to such result, which was shown in the Section \ref{section3.1}. Moreover, the fall of median ram pressure ($\rho v^2$) also showed similar trend with $\alpha_{\rm ram}$ = -2.54 (-2.53) for synthetic case and $\alpha_{\rm ram}$ = -2.09 (-2.03) for real case.

\textbf{Temperature.} We note that for CR2165 (CR2238), the median value of CME temperature decreases with time for real case with $\alpha_{\rm T}$ = -0.46 (-0.59), and synthetic case with $\alpha_{\rm T}$ = -1.48 (-1.42). The greater decrease in temperature (panel f1) is also associated with the lower decline in speed (panel d1) and can be attributed to the absence of SIRs and HSS in the uniform ambient through which the synthetic case CMEs propagate. As a result, the retarding force acting on the CME is weaker, allowing for more expansion and subsequent cooling compared to the real case CMEs. The disparity in the slope variation ($\alpha_{\rm T}$) between the real and synthetic cases suggests that the presence of higher density SIRs in the real case CMEs leads to compression, hindering the expansion and cooling process. This highlights that structured background solar wind with larger spatial gradient in density at different distances from the Sun plays an important role in the thermodynamic evolution of CMEs.

\subsection{CME Volume evolution}

    \begin{figure*}
       \centering
       \includegraphics[width = \textwidth]{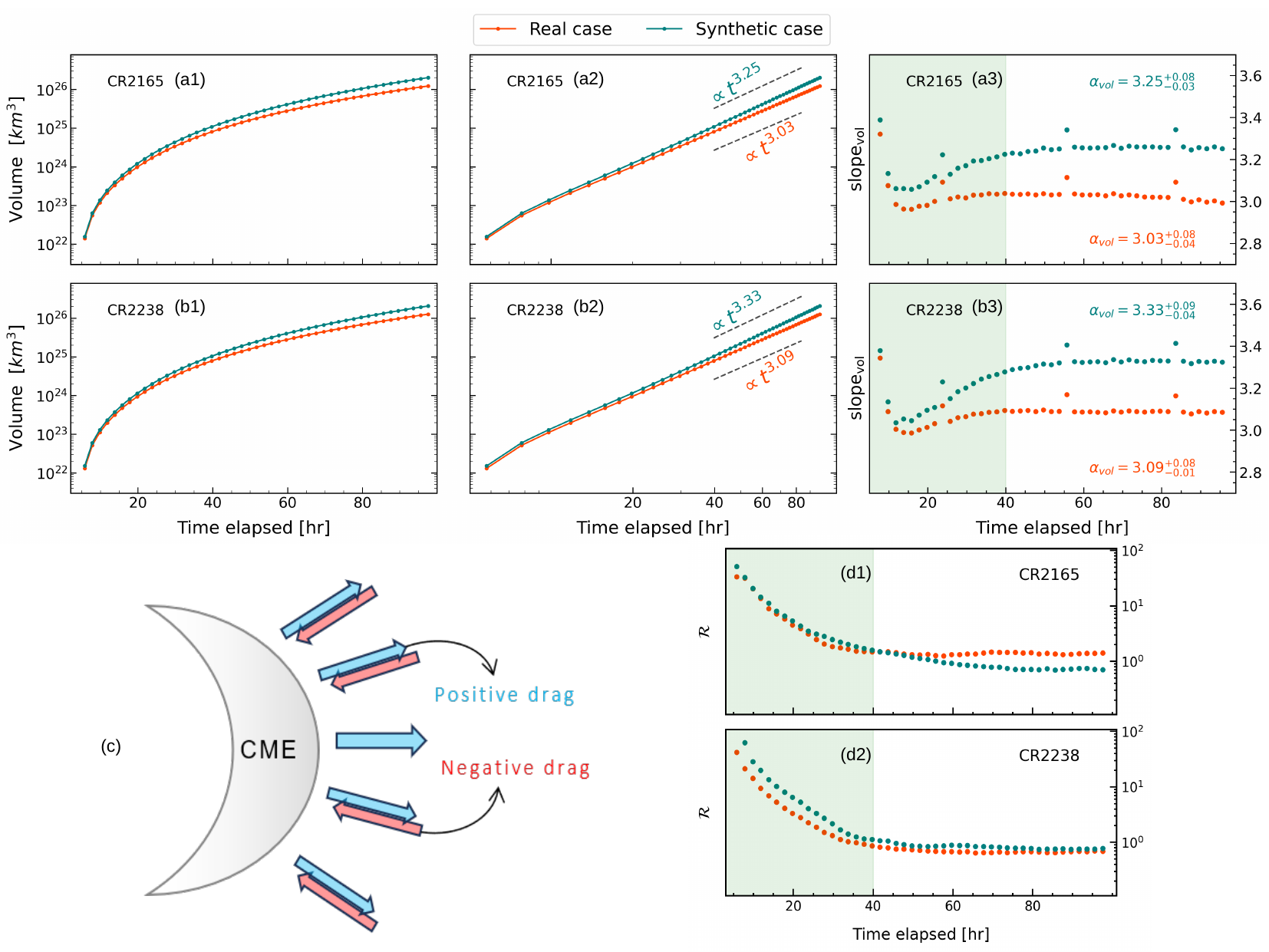}
       \caption{The temporal evolution of the total volume of the CME is presented for both the synthetic and real cases of CR2165 (a1, a2 and a3) and CR2238 (b1, b2 and b3). A rough sketch of negative and positive drag has been shown in subplot (c). And the evolution of the ratio of area on which positive and negative drag forces act is shown in the (d1) and (d2).}
       \label{fig:volume_evolution}
    \end{figure*}

The evolution of CME volume and its dependence on ambient SW conditions have not been thoroughly investigated in the past. The major challenge in such study, observational or numerical, is the absolute isolation of 3D CME structure from the background medium, specially at large distances from the Sun. In this regard, \cite{majumdar_2022_on} utilized the GCS model with observations from space and ground-based coronagraphs to study the 3D evolution of CME. However, their work focused on the total volume evolution of CME in the inner corona. In the inner-heliosphere, the density of CMEs is not high enough for unambiguous tracing through remote sensing observations, and thus a self-similar expansion is typically assumed to study the CME volume evolution. Previous studies, such as \cite{holzknecht_2019_cme} and \cite{temmer_2021_deriving}, employed the GCS fitting and assumed self-similar expansion to investigate CME evolution. 

In this work, we employed the structure identification technique detailed in Section \ref{section:CME_isolation} to examine the temporal evolution of total volume of CME and compare the real and synthetic cases to analyze the impact of ambient SW condition. Figure \ref{fig:volume_evolution} demonstrates the time evolution of CME volume for both real and synthetic cases, propagating in the ambient SW of CR2165, which occurred near solar maxima and CR2238, which was near solar minima. The computed values of total volume of CMEs are consistent with the values reported by \cite{holzknecht_2019_cme}. The rate of increase in the total volume of the synthetic case CME is found to be greater than that of the real case CME for both CRs. And as the time progresses, the disparity between the volume evolution of real and synthetic CMEs continues to grow until approximately 40 hours, after which it reaches a near-constant value, following a strict power law. From Figure \ref{fig:volume_evolution} subplot (a3) and (b3) it is evident that in the initial phase ($<$40 hours), the synthetic case CMEs exhibit a higher expansion rate compared to the real case CMEs. This discrepancy can be attributed to the presence of high density SIRs in the real case CMEs, which result in a greater accumulation of solar wind at the leading edge of the CME and the formation of a denser sheath region. This accumulation of sheath plasma could impede the expansion of the CME, leading to a slower increase in volume compared to the synthetic case CMEs. 

After the initial phase ($>$40 hours), the rate of increase in volume becomes nearly constant, the total volume of CME rises with $\alpha_{\rm vol}$ = 3.25 (3.32) for synthetic case and $\alpha_{\rm vol}$ = 3.03 (3.09) for real case of CR2165 (CR2238). It is worth noting that the strength of the SIR in CR2165 was greater than that in CR2238, and the $\alpha_{\rm vol}$ value for the CR2165 real case was lower than that in CR2238. As previously noted, synthetic case CMEs tend to have higher $\alpha_{\rm vol}$ values due to the absence of SIR. Consequently, CMEs that interact with weaker ambient conditions are more likely to have higher $\alpha_{\rm vol}$ values as well.

From our analysis, we can conclude that the temporal evolution of total volume of CME follows a power law beyond a certain propagation time or heliocentric distance from the Sun. We find that all the CMEs, real and synthetic in both CR, were having the power law exponent between 3.03 and 3.32 depending on the state of the background SW. For stronger ambient (presence of SIRs and high anisotropic medium), the exponent's value is smaller while its value is larger for weaker state of the heliosphere. In other words, if a CME is propagating through a dense non-uniform ambient, its total volume will be lesser compared to when it is propagating through a relatively tenuous and uniform ambient.

To identify the underlying factor leading to the power law behavior over time, we examined the drag force acting on the CME's surface. Depending on the specific location and instance, either the CME front pushes solar wind (positive drag) or experiences a pull (negative drag) by the ambient solar wind. By considering the effective areas associated with positive and negative drag forces, we introduced a ratio denoted as $\mathcal{R}$, which represents the number of grid points corresponding to positive and negative drag forces.

The temporal evolution of $\mathcal{R}$ exhibits a similar pattern to that of the volume evolution, characterized by two distinct phases and an asymptotic behavior after approximately 40 hours of propagation. The constant value of $\mathcal{R}$ indicates that the areas of positive and negative drag forces have reached a balanced state. It is important to note that the volume expansion of CMEs in this context is not self-similar. Instead, it is a dynamic process in which any changes are offset by compensating changes, resulting in an overall steady expansion rate. Additionally, as shown in Figure 9, the real case CMEs have achieved asymptotic behavior (in volume and $\mathcal{R}$ subplots) earlier than the ideal case CMEs. This indicates that higher the non-uniformity in the ambient SW, earlier the CME will achieve its steady state with surroundings.

Based on the above findings, we can conclude that the volume of a CME achieves a non-fractal power law state in relation to its surrounding environment after a specific duration of propagation in the inner heliosphere. The duration required to reach this state depends on the prevailing conditions of the ambient solar wind. Specifically, an isolated CME may attain the power law state in a shorter duration during the solar maxima phase, characterized by greater non-uniformity compared to the solar minima phase. These findings provide valuable insights into the intricate dynamics of CMEs and their interactions with the ambient solar wind. They emphasize the significant role played by drag forces in shaping the evolution of CME volume. In the subsequent section, we thoroughly explore the nature and magnitude of the drag force acting on CME.

\subsection{Drag force analysis}

\begin{figure*}
   \centering
   \includegraphics[width = \textwidth]{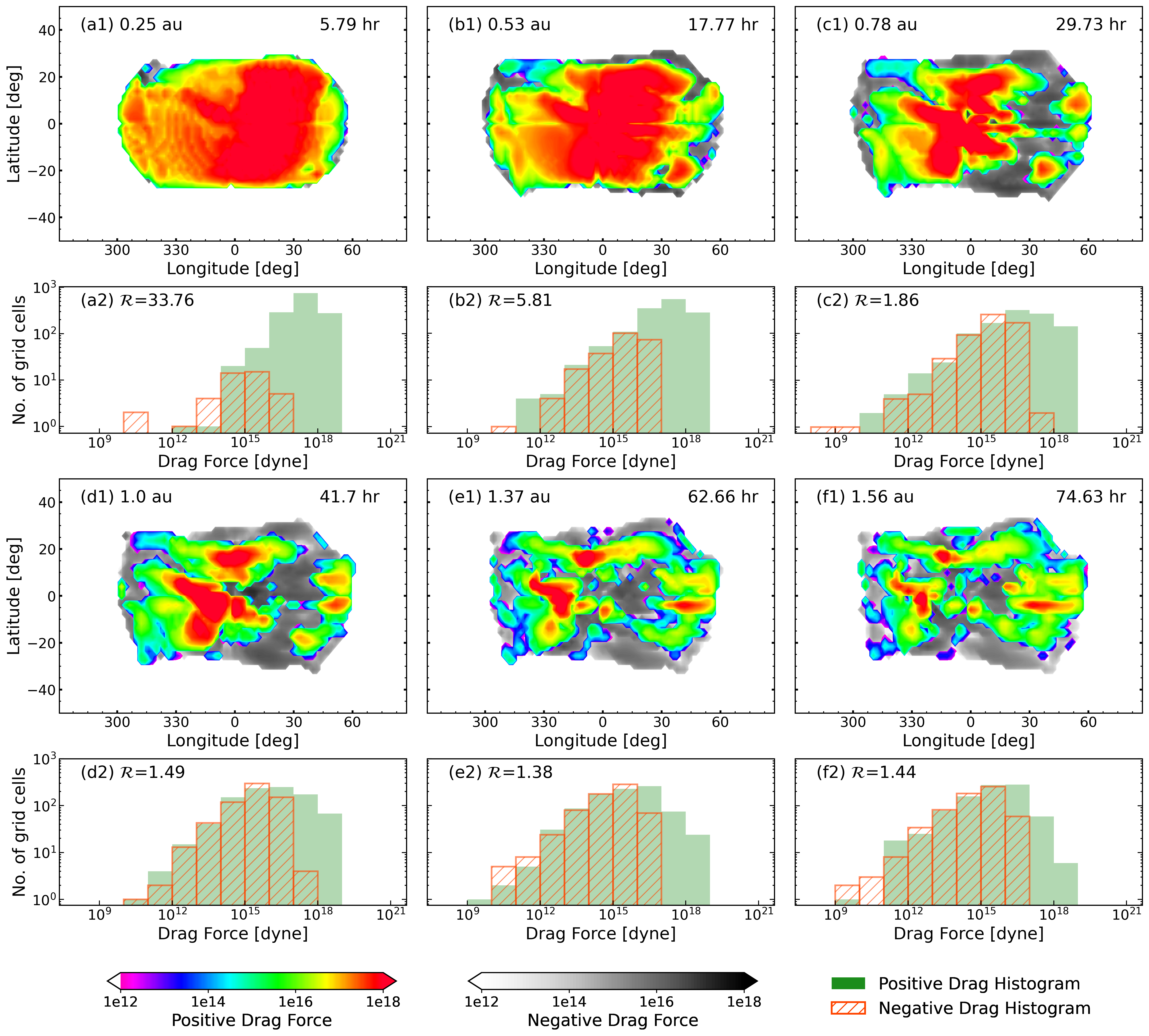}
   \caption{Figure showcases the drag force acting on the CME front due to anisotropic ambient conditions. Subplots (a1)-(f1) depicts the evolution of the drag force pattern on the front of real case CME of CR2165 as it advances in the heliosphere. The positive drag has been displayed using color map while, negative drag is in gray scale. Subplots (a2)-(f2) displays the histogram of magnitude of drag force corresponding to (a1)-(f1) subplots.}
   \label{fig:drag_force_pattern}
\end{figure*}

The deceleration of faster CMEs and the acceleration of slower CMEs during their propagation in the inner heliosphere are widely acknowledged phenomena \citep{gosling_1996_the, lindsay_1999_relationships, manoharan_2006_evolution}. Beyond a distance of 15 R$_\odot$, the dominant force governing CME dynamics is the aerodynamic drag force due to momentum exchange between CMEs and surrounding solar wind \citep{bojanvrnak_2008_the, sachdeva_2015_cme, kay_2021_modeling}. Consequently, several drag-based investigations have been conducted to study the dynamics of CMEs and their arrival time at the Earth \citep{mishra_2013_estimating, napoletano_2018_a, matejadumbovi_2021_dragbased}. While some research has focused on estimating the structural deformation of the CME front caused by drag, the nature of drag-force variation on the CME front remains unclear. In this study, using the SWASTi-CME model, we attempt to understand the evolution of this force on the CME front and its impact on CME deformation.

In this work, the drag-force ($\mathcal{F}_{\rm drag}$) between CME front and SW has been defined as:

\begin{equation} \label{eq:dragforce}
    \mathcal{F}_{\rm drag} = \frac{1}{2} \,\, C_d \,\, A \, \rho_{\rm sw}\, |v_{\rm CME} - v_{\rm sw}| (v_{\rm CME} - v_{\rm sw}) 
\end{equation}

where, $C_d$ is dimensionless drag-coefficient, $A$ is the area of contact between CME front and SW, $\rho_{\rm sw}$ is density of the ambient solar wind, $v_{\rm CME}$ and $v_{\rm sw}$ are velocity of CME and ambient SW, respectively \citep{bojanvrnak_2013_propagation,sachdeva_2015_cme}. A positive value of $\mathcal{F}_{\rm drag}$ indicates that the CME is pushing the ambient medium, while a negative drag implies that the ambient medium is pulling CME because of speed difference. The coefficient $C_d$ represents the level of interaction strength between the CME and the ambient medium, typically assumed to be of the order of unity. In this study, we have made the assumption that the drag coefficient remains constant at different heliocentric distances and set its value to one. The motivation is to explore the influence of density and relative velocity on the temporal evolution of the drag force acting on the CME.

The drag force on the CME front was calculated using Equation \ref{eq:dragforce}. To identify the entire CME front, we employed the CME isolation technique described in Section \ref{section:CME_isolation}. Subsequently, we examined the temporal evolution and distribution of the computed drag force. Figure \ref{fig:drag_force_pattern} illustrates the evolution of the drag force and its distribution on the CME front. In subplot \ref{fig:drag_force_pattern}(a1), the CME starts interacting with the ambient SW. We noticed that at a distance of 0.25 AU, the first SIR intercepts the CME's trajectory, leaving its structure imprinted in the drag force profile on the CME front. Furthermore, ripple-like structures are observed on the left side in subplot \ref{fig:drag_force_pattern} (a1), indicating the ploughing effect on the SW as the CME exerts drag force on the denser SIR. This implies that near the CME-SIR interaction region, the ambient SW exhibits negative meridional velocity, suggesting a counterflow of some SW plasma against the overall flow. As the CME progresses radially in the inner heliosphere, the negative drag becomes increasingly significant. Regions on the CME front where negative drag is observed indicate that the solar wind speed in those regions exceeds that of the CME. In other words, the presence of negative drag in that region causes the CME to undergo deformation and be pulled forward in relation to the other regions of the CME front. Furthermore, the imprint pattern resulting from the interaction with the SIR is observed to gradually diminish and shift towards the left. This observation aligns with the findings depicted in the subplots of Figure \ref{fig:4-step_evolution}, where the densest region formed in front of the leading edge is observed to shift eastward over time. And also the regions exhibiting negative drag force correspond to the deformation of the leading edge as illustrated in Figure \ref{fig:4-step_evolution}.

The presence of negative drag is predominantly observed on the eastward flank of the CME, while it is less pronounced on the westward flank. Initially, due to the higher ram pressure of the CME compared to the ambient solar wind, the CME was able to plow through the ambient medium and the first SIR without significant deformation. However, as the CME progresses and interacts with the second SIR, its effective ram pressure has significantly reduced due to earlier propagation, expansion, and drag. This reduction in ram pressure allows the SIR drag force to become negative, resulting in the deformation of the CME front. The decrease in ram pressure can be observed in Figure \ref{fig:median_evolution_2165} (c1), where it follows a power law with an exponent of -2.09. This indicates that the deformation of the CME front due to the drag force increases with time and radial distance, illustrating the progressive nature of the deformation process.

The histogram subplots presented in Figure \ref{fig:drag_force_pattern} illustrate the distribution of positive and negative drag forces acting on the CME front. Both positive and negative drag forces display a left-skewed normal distribution, where the median is greater than the mean. Initially, the distribution of positive drag force exhibits high skewness, which gradually decreases as the CME evolves in the heliosphere. This indicates that the area associated with positive drag decreases over time, leading to a reduction in the skewness of its distribution.

On the other hand, the distribution of negative drag force starts with low skewness but becomes more skewed as time passes. It is interesting to note that the area associated with negative drag expands rapidly and eventually reaches a point, at roughly 40 hours, where the ratio of the positive and negative drag areas ($\mathcal{R}$) becomes constant. In Figure \ref{fig:drag_force_pattern}, the magnitude and the distribution of grid points for both negative and positive drag forces almost coincide in the d2, e2, and f2 histogram plots. This indicates that not only have their effective areas reached a balanced state with each other, but their magnitudes have also stabilized. This indicates that the CME and the SW achieves a balanced state after approximately 40 hours of propagation.This potentially explains the observed power law expansion of CMEs discussed in the previous section.

Conclusively, the dynamic interaction between CME and SW leads to inter-play between positive and negative drag forces, which in turn, results in the localized deformation of CME front. This deformation is highly dependent on the presence of SIR and its 3D structure. While both forces reach a balanced state over time, the combined drag force decreases but remains positive throughout the simulation of the undertaken cases. Therefore, the nature of drag force directly affects the morphology and kinematics of CME in the inner-heliosphere and indirectly affects the distribution of CME's thermodynamic properties and time of arrival.

\section{Summary and Discussion} \label{section6}
In this work, we presented the newly developed MHD-based CME model, SWASTi-CME. It incorporates two distinct representations of CMEs: the elliptic cone model and the flux rope model. The elliptic cone model portrays a non-magnetic cloud with a simplified geometric shape. In addition to considering the angular width, this model also includes the angular height, leading to a better representation of the CME structure. On the other hand, the flux rope model allows for the inclusion of all the major deformities of the CME structure in the corona. The description of the initiation mechanisms for the cone and flux rope models into the MHD domain at 0.1 AU is also provided. Once the initial properties of the CMEs are defined, they undergo propagation in conjunction with the ambient solar wind generated by the SWASTi-SW module \citep{mayank_2022_swastisw}. 

Using SWASTi-CME, we conducted a simulation case study for CR2165, corresponding to solar maxima and consisting of five CMEs, and CR2238, corresponding to solar minima and consisting of one CME. Within this case study, the SWASTi results were also compared with the OMNI observational data at the Sun-Earth L1 point. The evaluation demonstrates that the results achieved for the flux rope CMEs exhibit favorable agreement with the observations for both scenarios: multiple (CR2165) and single (CR2238) CME events. However, while the cone CMEs yield satisfactory results for CR2238, their performance is comparatively less impressive for CR2165. This discrepancy in results for CR2165 can be attributed to the assumption of homogeneous speed and the absence of tilt in the CME axis, as well as the simpler insertion technique in the cone model. One way to enhance the accuracy of the cone CME results would be to take a distribution of CME speed as compared to a homogeneous value along the CME structure and consider the tilt angle. Additionally, improving the insertion mechanism beyond the one described in equation \ref{eq:half_time_width} may further improve the results.

In addition to the CME results, the simulated properties of the ambient solar wind also demonstrated a good agreement with the observations. It is worth noting that beyond 0.3 AU, the CME transitions into an incoherent MHD structure \citep{owens_2017_coronal}, highlighting the significance of accurately reconstructing both the ambient solar wind and the CME for a comprehensive assessment of CME dynamics. Considering the favorable performance of the flux rope CME and the ambient solar wind in comparison to the observations, it is reasonable to conclude that the SWASTi-CME model can be effectively utilized to investigate the interaction between a CME and ambient solar wind.

With the objective to investigate the influence of ambient solar wind conditions on CME evolution, we presented a setup of two cases: the \textit{real} case and the \textit{synthetic} case. The degree of uniformity in solar wind conditions is very high in synthetic case whereas, real case is the data-driven simulated background i.e., default run of SWASTi-CME. Both cases are described in detail in Section \ref{section4}, including the CME tracing technique employed to isolate the complete 3D structure of the CME from the ambient solar wind. Using this setup of two cases, we presented an analysis on the impact of non-uniformity in the ambient solar wind conditions on CMEs during the CR2165 (near solar maxima) and CR2238 (near solar minima) periods, particularly focusing on HSSs and SIRs. This analysis of the interaction between the CME and solar wind has yielded several findings, which are discussed as follows:

\begin{enumerate}[label={\tiny \textbullet}, leftmargin=*, itemindent=+0.75em]
  \item \textit{Morphology}: An initial eastward movement of the CME was observed, indicating a deflection caused by the obstructive effect of spiral-shaped ambient solar wind streams propagating from west to east. The trajectory of the CME can be further influenced by the presence of high-density SIRs, resulting in a deflection along the streamline of the SIR and causing an eastward deviation. Furthermore, on investigating the expansion of CMEs, we noticed distinct behaviors between the CME flank within a HSS and the flank interacting with the SIR. The former underwent over-expansion, while the latter remained under-expanded. These variations in expansion rates have the potential to induce changes in the density distribution of the CMEs. Additionally, the spatial variation in density, which is correlated with the solar wind speed and, consequently, the drag force, along the leading edges of the CMEs contributes to their deformation. Notably, in the presence of SIRs, characterized by denser regions with significant speed variations, this phenomenon becomes more pronounced. As a result of the diverse drag forces experienced by different regions of the CME front, some regions undergo radial displacement, being pushed rearward, while others experience slight advancement due to over-expansion.
  
  \item \textit{Internal pressure}: The analysis of temporal evolution reveals notable trends in CME pressure distribution, including thermal and magnetic pressure. The findings indicate that an uniform background (synthetic case) leads to a more pronounced decline in CME pressure compared to a non-uniform background (real case). This suggests that higher anisotropy in the solar wind leads to higher pressure within the CME at 1 AU. Moreover, this disparity is particularly prominent during the early stages of CME propagation, specifically within the first 24 hours after crossing 21.5 R$_\odot$. Subsequently, the thermal pressure follows a power law decay (though not strictly) with a slope of -2.30$^{+0.54}_{-0.37}$ (-2.50$^{+0.23}_{-0.1}$), while the magnetic pressure demonstrates a similar power law decay with a slope of -2.37$^{+0.51}_{-0.24}$ (-2.24$^{+0.31}_{-0.28}$) for real case of CR2165 (CR2238). These findings provide valuable insights into the influence of anisotropy in the ambient solar wind on the pressure dynamics within CMEs.

  \item \textit{Internal speed}: The distribution of CME speed displayed a gradual decrease over time. In the synthetic case, representing an isotropic background, the power law exponent was -0.02$^{+0.01}_{-0.01}$ (-0.02$^{+0.02}_{-0.02}$), while in the real case, representing an anisotropic background, of CR2165 (CR2238), the power law exponent was -0.1$^{+0.03}_{-0.02}$ (-0.06$^{+0.01}_{-0.01}$). Notably, the median speed of both CR2165 and CR2238 CMEs showed a more pronounced decline in the anisotropic solar wind conditions compared to the isotropic conditions, which can be attributed to increased drag resulting from their interaction with the ambient solar wind. Furthermore, the speed reduction is more significant in CR2165 due to the presence of stronger SIRs along its path, leading to a more substantial deceleration effect. These findings emphasize the influence of solar wind anisotropy on CME dynamics, highlighting the role of SIRs and drag force in shaping CME speed profiles.

  \item \textit{Internal density and temperature}: While both the real and synthetic case CMEs experience a decrease in median proton density over time, they display distinct patterns in the temporal evolution of the power law exponent observed for CR2165 and CR2238. This highlights the influence of ambient solar wind conditions on the distribution of CME density throughout their propagation in the simulation domain. Furthermore, the temperature of CMEs gradually decreases, with the CME in an isotropic solar wind demonstrating a significantly higher rate of decrement. The difference in the temporal evolution of the power law exponent between the anisotropic and isotropic backgrounds indicates that the cooling process may encounter hindrance due to the interaction with the ambient solar wind. This hindrance can be attributed to the CME encountering higher-density SIRs, leading to compression and inhibition of the CME's expansion and cooling process.
  
  \item \textit{Total Volume}: The analysis of the temporal evolution of the total volume of CMEs revealed that CMEs expand at a greater rate in conditions of higher isotropy in the ambient solar wind. The presence of high-density SIRs in the real case CMEs led to the formation of a denser sheath region at the leading edge, impeding their expansion and resulting in a slower increase in volume compared to the synthetic case CMEs. This difference in expansion rates between the isotropic and anisotropic backgrounds increased over time until approximately 40 hours since entering the MHD domain. After this time, the expansion rates reached a near-constant value, following a strict power law with different exponents. Similar, two phase expansion of CME was also observed by \cite{scolini_2022_causes}. The real case CMEs exhibited exponent values of 3.03 for CR2165, which had stronger SIRs, and 3.09 for CR2238, which had weaker SIRs. The stronger ambient conditions, i.e. the presence of SIRs and highly anisotropic medium, resulted in lower power law exponents and consequently smaller total volumes. 
  
  Though the results indicate that total volume varies roughly in proportion to the cube of time, the expansion is not self-similar in the heliosphere. Previous studies have suggested that most CMEs tend to expand approximately self-similarly in the corona \citep{vourlidas_2010_comprehensive, subramanian_2014_selfsimilar}. However, a recent study by \cite{carlosrobertobraga_2022_coronal}, based on WISPR observations, demonstrated that while CMEs expand self-similarly in the corona, they start deforming at ~0.1 AU due to the non-uniformity of solar wind speed. Similarly, findings from our simulation also align with this argument, showing that the deformation of the CME deviates its structure from self-similar expansion.

  \item \textit{Drag Force}: The investigation into the temporal evolution and distribution of the drag force acting on the CME front revealed interesting findings. The interaction between the CME and SIR resulted in the imprinting of the SIR's structure on the CME front, leading to higher drag force values on CME font. The presence of ripple-like structures indicated the dominant ploughing effect during the initial hours of CME-SW interaction. Furthermore, the anisotropic nature of the ambient solar wind conditions resulted in an uneven distribution of drag force on the CME front, with some regions experiencing positive force while others experienced negative force. This resulted in an asymmetric deformation of the CME's leading edge. The distribution of drag force exhibited a left-skewed normal distribution for both negative and positive values. Although the negative drag force grew rapidly as the CME evolved in the heliosphere, it consistently remained approximately 10 times smaller than the positive drag force, resulting in a decelerating forward motion of the CME.
  
\end{enumerate}

In summary, our study has primarily focused on investigating the structural and thermodynamic effects of the ambient solar wind on CMEs in the heliosphere. It is important to emphasize that such changes in CME properties can also arise from interactions with other CMEs. The adopted distribution of magnetic field within the FRi3D model in the CME has been validated with in-situ signatures of CME. However, a comparison with the other common models like Lundquist \citep{lundquist1950magnetohydrostatic} and Gold and Hoyle \citep{gold_1960_on} would be a natural future step. Moreover, employing a cylindrical geometry to populate curved flux rope may lead to imbalance between positive and negative flux. A curved magnetic flux rope model, such as proposed by \cite{singh_2022_ensemble} in which they used the toroidal flux model of \cite{mvandas_2017_magnetic}, could address this issue. Interestingly, a recent investigation carried out by \cite{davies_2022_multispacecraft} has demonstrated a decreasing orientation of the flux rope over time as the CME propagates. These intriguing aspects will be the central focus of our future endeavors, as we plan to conduct a dedicated analysis on the magnetic field profile of the flux rope and explore the interactions between multiple CMEs using the SWASTi-CME model.

\begin{acknowledgments}
PM extends his gratitude to Anwesha Maharana for her invaluable discussions concerning the FRi3D model. PM gratefully acknowledges the support provided by the Prime Minister's Research Fellowship. BV and DC acknowledge the support from the ISRO RESPOND grant number: ISRO/RES/2/436/21-22. Furthermore, the work of DC is supported by the Department of Space, Government of India.
\par
The used GONG synoptic magnetograms maps can be freely obtained from \url{https://gong.nso.edu/data/magmap/crmap.html}. The OMNI data are taken from the Goddard Space Flight Center, accessible at \url{https://spdf.gsfc.nasa.gov/pub/data/omni/}. The PLUTO code used for MHD simulation can be downloaded free of charge from \url{http://plutocode.ph.unito.it/}.
\par
The data-set presented in this paper is archived at \url{https://doi.org/10.5281/zenodo.10011252}. The archived repository contains the data and python script used for the model validation with observations. The data pertaining to other specific Carrington Rotations from potential users will be shared on reasonable request to the corresponding author.
\end{acknowledgments}





\appendix

\section{Magnetic field Orientation in FRi3D} \label{Appendix1}

    \begin{figure*}
       \centering
       \includegraphics[width = \textwidth]{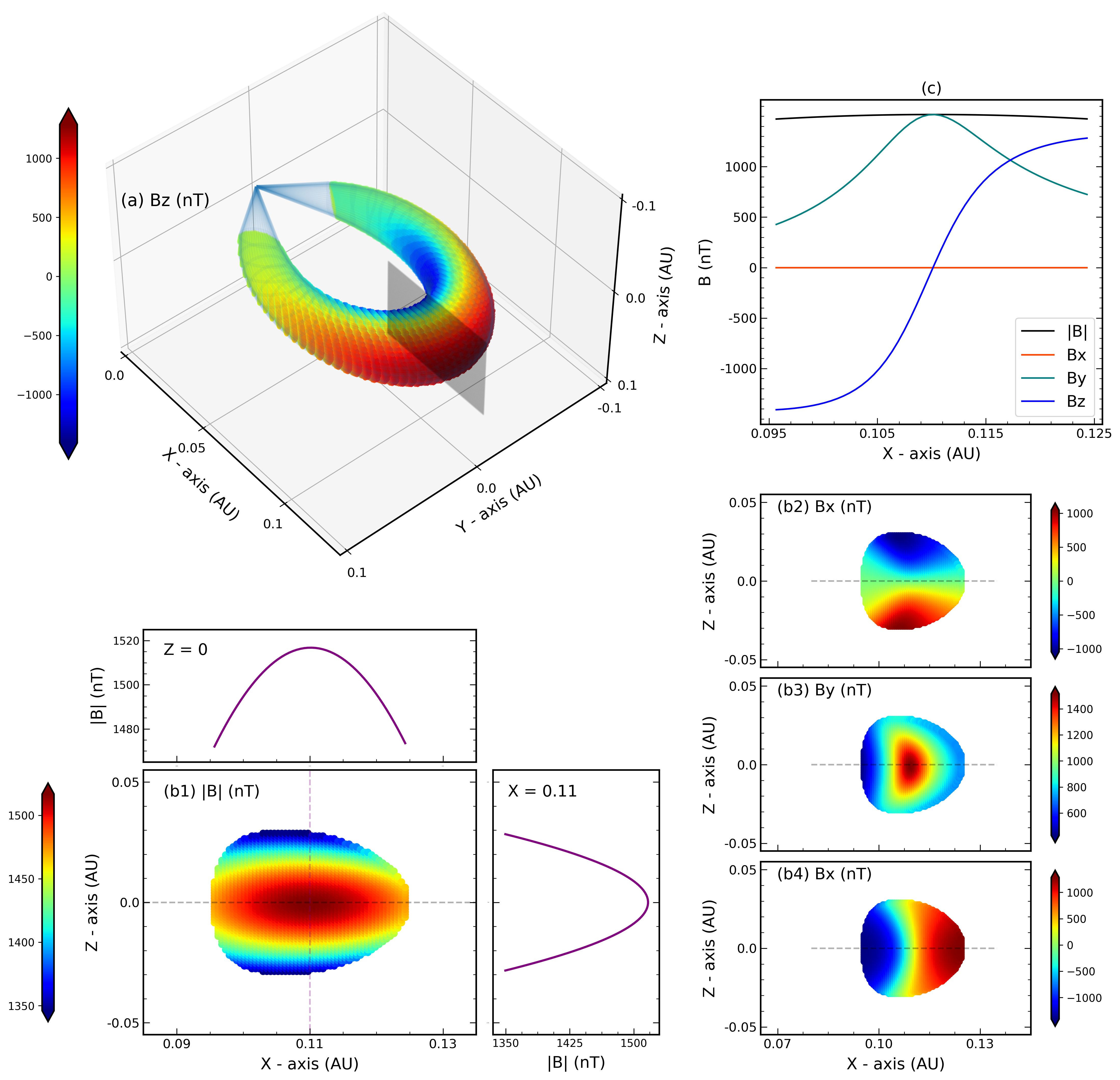}
       \caption{The picture depicts the magnetic profile of CME formed through FRi3D, where the strength of magnetic fields are color coded. The 3D profile of Bz is shown in subplot (a), whose 2D slice has been cut in X-Z plan at Y=0 and shown in subplots b1-b4. The 1D profile of magnetic field at Z=0 of this 2D slice is shown in subplot (c).}
       \label{fig:FRi3D_mag_orientation}
    \end{figure*}

In this section, we demonstrate the magnetic profile of Flux Rope in 3D \cite[FRi3D,][]{isavnin_2016_fried} CME. The FRi3D model is an open-source python package (\url{https://pypi.org/project/ai.fri3d/}) that offers an analytical 3D model of flux rope CME. It incorporates major deformations of the structure and is capable in reproducing the global geometric and magnetic properties  of a CME. In current version of FRi3D model, the strength of the flux rope is estimated using a bivariate normal distribution with constant twist \citep{maharana_2023_updates} and has been implemented in multiple studies \citep{maharana_2022_implementation, maharana_2023_rotation, palmerio_2023_modeling}. \\

Figure \ref{fig:FRi3D_mag_orientation} illustrates the typical magnetic field structure in the FRi3D CME. Subplot (a) displays the 3D profile of B$_{\rm z}$ of the CME, generated using the following parameter values: $\theta_{\rm CME} = 0$\textdegree{}, $\phi_{\rm CME} = 0$\textdegree{}, $\varphi_{\rm hw} = 51$\textdegree{}, $\varphi_{\rm hh} = 15$\textdegree{}, $R_{\rm t} = 0.11$ AU, $\gamma = 0$\textdegree{}, $\varphi_{\rm p} = 0.5, \eta = 0.4, \phi_{\rm B} = 1e13$ Wb and $\tau = 2.0$. The definitions of these parameters are given in Table \ref{tab:tabel1}. To display the orientation of magnetic fields inside the flux rope, a 2D slice of the X-Z plane at Y = 0 has been shown in subplots (b1-b4). The bivariate normal distribution of total magnetic field strength can be seen in the subplot (b1). The magnitude variation along the center of the cross-section at Z=0 and X=0.11 is also shown in its sub set plots, which is form a bell-shaped structure. Additionally, the 1D profile along the X-axis at Z = 0, as marked by the black dashed horizontal line, of this slice is also presented in subplot (c), which closely resembles with the observed in-situ signatures of flux rope CME. \\

To further validate the existing analytical form of the magnetic field profile, we compared the properties of a CME constructed using the FRi3D model with OMNI data at the Sun-Earth L1 point. We selected a CME that reached spacecraft on 2 January 2010 and demonstrated a smooth rotation in its magnetic field. We fine-tuned the FRi3D parameters to achieve the best-fit with the in-situ OMNI data. For simplicity, we constructed the interplanetary CME and estimated the 1D profile at spacecraft's location, using an approach similar to that in Figure \ref{fig:FRi3D_mag_orientation}(c). \\

\begin{figure*}
       \centering
       \includegraphics[width = 0.54\textwidth]{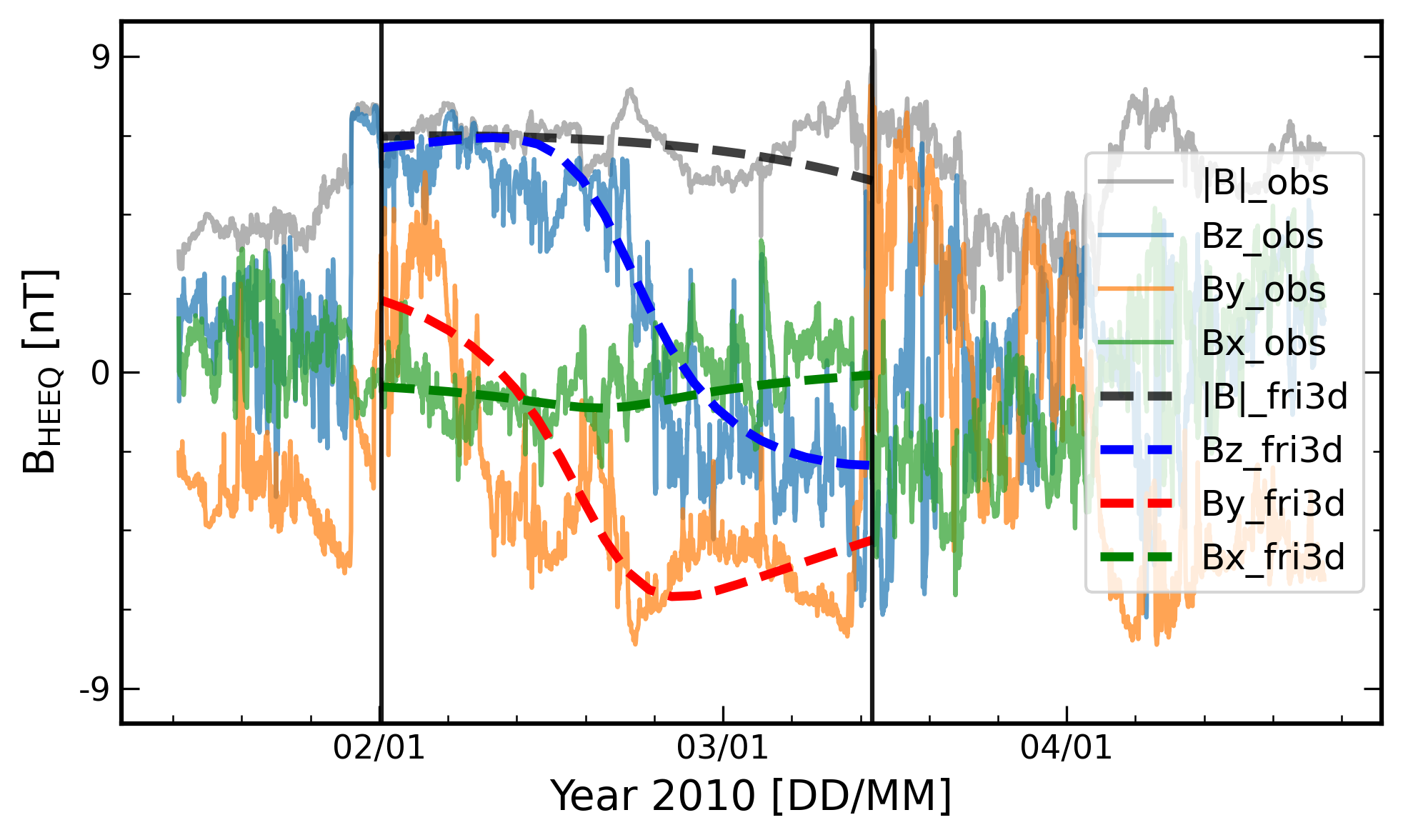}
       \caption{Comparison of FRi3D CME with the observed OMNI data having in-situ signatures of CME. The magnetic field components are in nT and HEEQ coordinate system.}
       \label{fig:FRi3D_OMNI}
    \end{figure*}

Figure \ref{fig:FRi3D_OMNI} presents the comparison of the fine-tuned CME parameter values to the in-situ OMNI data of January 2010. All the values are displayed in the Heliocentric Earth Equatorial (HEEQ) coordinate system and the following parameter values were used: $\theta_{\rm CME} = -4$\textdegree{}, $\phi_{\rm CME} = -9$\textdegree{}, $\varphi_{\rm hw} = 40$\textdegree{}, $\varphi_{\rm hh} = 18$\textdegree{}, $R_{\rm t} = 1.0$ AU, $\gamma = -48$\textdegree{}, $\varphi_{\rm p} = 0.4, \eta = 0.6, \phi_{\rm B} = 2e12$ Wb, $\tau = 1.7$, chirality = -1 and polarity = -1. The magnetic field strength constructed using the FRi3D model aligns well with observed values, and the nature of variation in all three components (B$_{\rm x}$, B$_{\rm y}$ and B$_{\rm z}$) are also consistent with the in-situ data.

\bibliography{SWASTi_CME1}{}

\begin{thebibliography}{}
\expandafter\ifx\csname natexlab\endcsname\relax\def\natexlab#1{#1}\fi
\providecommand{\url}[1]{\href{#1}{#1}}
\providecommand{\dodoi}[1]{doi:~\href{http://doi.org/#1}{\nolinkurl{#1}}}
\providecommand{\doeprint}[1]{\href{http://ascl.net/#1}{\nolinkurl{http://ascl.net/#1}}}
\providecommand{\doarXiv}[1]{\href{https://arxiv.org/abs/#1}{\nolinkurl{https://arxiv.org/abs/#1}}}

\bibitem[{Arge(2003)}]{arge_2003_improved}
Arge, C.~N. 2003, AIP Conference Proceedings, 679, 190, \dodoi{10.1063/1.1618574}

\bibitem[{Biondo {et~al.}(2021)Biondo, Pagano, Reale, \& Bemporad}]{biondo_2021_tracing}
Biondo, R., Pagano, P., Reale, F., \& Bemporad, A. 2021, Astronomy and Astrophysics, 654, L3, \dodoi{10.1051/0004-6361/202141892}

\bibitem[{Braga {et~al.}(2022)Braga, Vourlidas, Liewer, Hess, Stenborg, \& Riley}]{carlosrobertobraga_2022_coronal}
Braga, C.~R., Vourlidas, A., Liewer, P.~C., {et~al.} 2022, The Astrophysical Journal, 938, 13, \dodoi{10.3847/1538-4357/ac90bf}

\bibitem[{Davies {et~al.}(2022)Davies, Winslow, Scolini, Forsyth, Möstl, Lugaz, \& Galvin}]{davies_2022_multispacecraft}
Davies, E.~L., Winslow, R.~M., Scolini, C., {et~al.} 2022, The Astrophysical Journal, 933, 127, \dodoi{10.3847/1538-4357/ac731a}

\bibitem[{Dumbović {et~al.}(2021)Dumbović, Čalogović, Martinić, Vršnak, Sudar, Temmer, \& Veronig}]{matejadumbovi_2021_dragbased}
Dumbović, M., Čalogović, J., Martinić, K., {et~al.} 2021, Frontiers in Astronomy and Space Sciences, 8, \dodoi{10.3389/fspas.2021.639986}

\bibitem[{Dumbović {et~al.}(2018)Dumbović, Čalogović, Vršnak, Temmer, Mays, Veronig, \& Piantschitsch}]{matejadumbovi_2018_the}
Dumbović, M., Čalogović, J., Vršnak, B., {et~al.} 2018, The Astrophysical Journal, 854, 180, \dodoi{10.3847/1538-4357/aaaa66}

\bibitem[{Geyer {et~al.}(2023)Geyer, Dumbović, Temmer, Veronig, Dissauer, \& Vršnak}]{geyer_2023_interaction}
Geyer, P., Dumbović, M., Temmer, M., {et~al.} 2023, Astronomy and Astrophysics, 672, A168, \dodoi{10.1051/0004-6361/202245433}

\bibitem[{Gibson \& Low(1998)}]{gibson_1998_a}
Gibson, S.~E., \& Low, B.~C. 1998, The Astrophysical Journal, 493, 460, \dodoi{10.1086/305107}

\bibitem[{Gold \& Hoyle(1960)}]{gold_1960_on}
Gold, T., \& Hoyle, F. 1960, Monthly Notices of the Royal Astronomical Society, 120, 89, \dodoi{10.1093/mnras/120.2.89}

\bibitem[{Gosling \& Pizzo(1999)}]{gosling_1999_formation}
Gosling, J.~T., \& Pizzo, V.~J. 1999, Space Science Reviews, 89, 21, \dodoi{10.1007/978-94-017-1179-1_3}

\bibitem[{Gosling \& Riley(1996)}]{gosling_1996_the}
Gosling, J.~T., \& Riley, P. 1996, Geophysical Research Letters, 23, 2867, \dodoi{10.1029/96gl02843}

\bibitem[{Harrison {et~al.}(2012)Harrison, Davies, Möstl, Liu, Temmer, Bisi, Eastwood, de, Nitta, Amerstorfer, Farrugia, Forsyth, Jackson, ~, Kilpua, Odstrcil, \& Webb}]{harrison_2012_an}
Harrison, R.~M., Davies, J.~C., Möstl, C., {et~al.} 2012, The Astrophysical Journal, 750, 45, \dodoi{10.1088/0004-637x/750/1/45}

\bibitem[{Heinemann {et~al.}(2019)Heinemann, Temmer, Farrugia, Dissauer, Kay, Wiegelmann, Dumbović, Veronig, Podladchikova, Hofmeister, Lugaz, \& Carcaboso}]{heinemann_2019_cmehss}
Heinemann, S.~G., Temmer, M., Farrugia, C.~J., {et~al.} 2019, Solar Physics, 294, \dodoi{10.1007/s11207-019-1515-6}

\bibitem[{Hinterreiter {et~al.}(2021)Hinterreiter, Amerstorfer, Temmer, Reiss, Weiss, Möstl, Barnard, Pomoell, Bauer, \& Amerstorfer}]{jrgenhinterreiter_2021_dragbased}
Hinterreiter, J., Amerstorfer, T., Temmer, M., {et~al.} 2021, Space Weather, 19, \dodoi{10.1029/2021sw002836}

\bibitem[{Holzknecht {et~al.}(2019)Holzknecht, Temmer, Dumbovic, Wellenzohn, Krikova, Heinemann, Rodari, Vrsnak, \& Veronig}]{holzknecht_2019_cme}
Holzknecht, L., Temmer, M., Dumbovic, M., {et~al.} 2019, arXiv:1904.11418 [astro-ph, physics:physics], \dodoi{10.48550/arXiv.1904.11418}

\bibitem[{Hu {et~al.}(2015)Hu, Qiu, \& Krucker}]{hu_2015_magnetic}
Hu, Q., Qiu, J., \& Krucker, S. 2015, Journal of Geophysical Research: Space Physics, 120, 5266, \dodoi{10.1002/2015ja021133}

\bibitem[{Isavnin(2016)}]{isavnin_2016_fried}
Isavnin, A. 2016, The Astrophysical Journal, 833, 267, \dodoi{10.3847/1538-4357/833/2/267}

\bibitem[{Janvier {et~al.}(2019)Janvier, Winslow, Good, Bonhomme, Démoulin, Dasso, Möstl, Lugaz, Amerstorfer, Soubrié, \& Boakes}]{janvier_2019_generic}
Janvier, M., Winslow, R.~M., Good, S., {et~al.} 2019, Journal of Geophysical Research: Space Physics, 124, 812, \dodoi{10.1029/2018ja025949}

\bibitem[{Jin {et~al.}(2017)Jin, Manchester, ~, Sokolov, Tóth, Mullinix, Taktakishvili, Chulaki, \& Gombosi}]{jin_2017_dataconstrained}
Jin, M., Manchester, W.~B., ~, v., {et~al.} 2017, The Astrophysical Journal, 834, 173, \dodoi{10.3847/1538-4357/834/2/173}

\bibitem[{Kay \& Nieves-Chinchilla(2021)}]{kay_2021_modeling}
Kay, C., \& Nieves-Chinchilla, T. 2021, Journal of Geophysical Research: Space Physics, 126, \dodoi{10.1029/2020ja028911}

\bibitem[{Kay {et~al.}(2022)Kay, Nieves‐Chinchilla, Hofmeister, \& Palmerio}]{kay_2022_beyond}
Kay, C., Nieves‐Chinchilla, T., Hofmeister, S.~J., \& Palmerio, E. 2022, Space Weather, 20, \dodoi{10.1029/2022sw003165}

\bibitem[{Lindsay {et~al.}(1999)Lindsay, Luhmann, Russell, \& Gosling}]{lindsay_1999_relationships}
Lindsay, G.~M., Luhmann, J.~G., Russell, C.~T., \& Gosling, J.~T. 1999, Journal of Geophysical Research: Space Physics, 104, 12515, \dodoi{10.1029/1999ja900051}

\bibitem[{Liu {et~al.}(2005)Liu, Richardson, \& Belcher}]{liu_2005_a}
Liu, Y., Richardson, J., \& Belcher, J. 2005, Planetary and Space Science, 53, 3, \dodoi{10.1016/j.pss.2004.09.023}

\bibitem[{Liu {et~al.}(2019)Liu, Shen, \& Yang}]{liu_2019_numerical}
Liu, Y.-S., Shen, F., \& Yang, Y. 2019, The Astrophysical Journal, 887, 150, \dodoi{10.3847/1538-4357/ab543e}

\bibitem[{Liu {et~al.}(2012)Liu, Luhmann, Möstl, Martinez-Oliveros, Bale, Lin, Harrison, Temmer, Webb, \& Odstrcil}]{liu_2012_interactions}
Liu, Y.~W., Luhmann, J.~G., Möstl, C., {et~al.} 2012, The Astrophysical Journal Letters, 746, L15, \dodoi{10.1088/2041-8205/746/2/l15}

\bibitem[{Lundquist(1950)}]{lundquist1950magnetohydrostatic}
Lundquist, S. 1950, Ark. Fys., 2, 361

\bibitem[{Maharana {et~al.}(2022)Maharana, Isavnin, Scolini, Wijsen, Rodriguez, Mierla, Magdalenić, \& Poedts}]{maharana_2022_implementation}
Maharana, A., Isavnin, A., Scolini, C., {et~al.} 2022, Advances in Space Research, 70, 1641, \dodoi{10.1016/j.asr.2022.05.056}

\bibitem[{Maharana {et~al.}(2023{\natexlab{a}})Maharana, Plets, Isavnin, \& Poedts}]{maharana_2023_updates}
Maharana, A., Plets, K., Isavnin, A., \& Poedts, S. 2023{\natexlab{a}}, arXiv.org, \dodoi{10.48550/arXiv.2310.11402}

\bibitem[{Maharana {et~al.}(2023{\natexlab{b}})Maharana, Scolini, Schmieder, \& Poedts}]{maharana_2023_rotation}
Maharana, A., Scolini, C., Schmieder, B., \& Poedts, S. 2023{\natexlab{b}}, Astronomy and Astrophysics, 675, 136, \dodoi{10.1051/0004-6361/202345902}

\bibitem[{Majumdar {et~al.}(2022)Majumdar, Patel, \& Pant}]{majumdar_2022_on}
Majumdar, S., Patel, R., \& Pant, V. 2022, The Astrophysical Journal, 929, 11, \dodoi{10.3847/1538-4357/ac5909}

\bibitem[{Manchester {et~al.}(2017)Manchester, Kilpua, Liu, Lugaz, Riley, Török, \& Vršnak}]{manchester_2017_the}
Manchester, W., Kilpua, E. K.~J., Liu, Y.~D., {et~al.} 2017, Space Science Reviews, 212, 1159, \dodoi{10.1007/s11214-017-0394-0}

\bibitem[{Manoharan(2006)}]{manoharan_2006_evolution}
Manoharan, P.~K. 2006, Solar Physics, 235, 345, \dodoi{10.1007/s11207-006-0100-y}

\bibitem[{Mayank {et~al.}(2022)Mayank, Vaidya, \& Chakrabarty}]{mayank_2022_swastisw}
Mayank, P., Vaidya, B., \& Chakrabarty, D. 2022, The Astrophysical Journal Supplement Series, 262, 23, \dodoi{10.3847/1538-4365/ac8551}

\bibitem[{Mignone {et~al.}(2007)Mignone, Bodo, Massaglia, Matsakos, Tesileanu, Zanni, \& Ferrari}]{mignone_2007_pluto}
Mignone, A., Bodo, G., Massaglia, S., {et~al.} 2007, The Astrophysical Journal Supplement Series, 170, 228, \dodoi{10.1086/513316}

\bibitem[{Mishra {et~al.}(2021)Mishra, Doshi, \& Srivastava}]{mishra_2021_radial}
Mishra, W., Doshi, U., \& Srivastava, N. 2021, Frontiers in Astronomy and Space Sciences, 8, \dodoi{10.3389/fspas.2021.713999}

\bibitem[{Mishra \& Srivastava(2013)}]{mishra_2013_estimating}
Mishra, W., \& Srivastava, N. 2013, The Astrophysical Journal, 772, 70, \dodoi{10.1088/0004-637x/772/1/70}

\bibitem[{Mishra {et~al.}(2014)Mishra, Srivastava, \& Davies}]{mishra_2014_a}
Mishra, W., Srivastava, N., \& Davies, J.~A. 2014, The Astrophysical Journal, 784, 135, \dodoi{10.1088/0004-637x/784/2/135}

\bibitem[{Mishra {et~al.}(2017)Mishra, Wang, Srivastava, \& Shen}]{mishra_2017_assessing}
Mishra, W., Wang, Y., Srivastava, N., \& Shen, C. 2017, The Astrophysical Journal Supplement Series, 232, 5, \dodoi{10.3847/1538-4365/aa8139}

\bibitem[{Napoletano {et~al.}(2018)Napoletano, Forte, Moro, Pietropaolo, Giovannelli, \& Berrilli}]{napoletano_2018_a}
Napoletano, G., Forte, R., Moro, D.~D., {et~al.} 2018, Journal of Space Weather and Space Climate, 8, A11, \dodoi{10.1051/swsc/2018003}

\bibitem[{Odstrcil {et~al.}(2004)Odstrcil, Riley, \& Zhao}]{dusanodstrcil_2004_numerical}
Odstrcil, D., Riley, P., \& Zhao, X.~S. 2004, Journal of Geophysical Research, 109, \dodoi{10.1029/2003ja010135}

\bibitem[{Owens {et~al.}(2017)Owens, Lockwood, \& Barnard}]{owens_2017_coronal}
Owens, M.~J., Lockwood, M., \& Barnard, L.~A. 2017, Scientific Reports, 7, \dodoi{10.1038/s41598-017-04546-3}

\bibitem[{Palmerio {et~al.}(2023)Palmerio, Maharana, Lynch, Scolini, Good, Pomoell, Isavnin, \& Kilpua}]{palmerio_2023_modeling}
Palmerio, E., Maharana, A., Lynch, B.~J., {et~al.} 2023, arXiv (Cornell University), \dodoi{10.48550/arxiv.2310.05846}

\bibitem[{Papitashvili \& King(2020)}]{Papi2020}
Papitashvili, N.~E., \& King, J.~H. 2020, ``OMNI 5-min Data" [Data set], NASA Space Physics Data Facility, \dodoi{10.48322/gbpg-5r77}

\bibitem[{Pomoell \& Poedts(2018)}]{pomoell_2018_euhforia}
Pomoell, J., \& Poedts, S. 2018, Journal of Space Weather and Space Climate, 8, A35, \dodoi{10.1051/swsc/2018020}

\bibitem[{Raghav \& Shaikh(2020)}]{anilraghav_2020_the}
Raghav, A., \& Shaikh, Z.~A. 2020, Monthly Notices of the Royal Astronomical Society: Letters, 493, L16, \dodoi{10.1093/mnrasl/slz187}

\bibitem[{Richardson(2018)}]{richardson_2018_solar}
Richardson, I.~G. 2018, Living Reviews in Solar Physics, 15, \dodoi{10.1007/s41116-017-0011-z}

\bibitem[{Riley \& Crooker(2004)}]{riley_2004_kinematic}
Riley, P., \& Crooker, N.~U. 2004, The Astrophysical Journal, 600, 1035, \dodoi{10.1086/379974}

\bibitem[{Rollett {et~al.}(2016)Rollett, Möstl, Isavnin, Davies, Kubicka, Amerstorfer, \& Harrison}]{rollett_2016_elevohi}
Rollett, T., Möstl, C., Isavnin, A., {et~al.} 2016, The Astrophysical Journal, 824, 131, \dodoi{10.3847/0004-637x/824/2/131}

\bibitem[{Sachdeva {et~al.}(2015)Sachdeva, Subramanian, Colaninno, \& Vourlidas}]{sachdeva_2015_cme}
Sachdeva, N., Subramanian, P., Colaninno, R., \& Vourlidas, A. 2015, The Astrophysical Journal, 809, 158, \dodoi{10.1088/0004-637x/809/2/158}

\bibitem[{Salman {et~al.}(2020)Salman, Winslow, \& Lugaz}]{salman_2020_radial}
Salman, T.~M., Winslow, R.~M., \& Lugaz, N. 2020, Journal of Geophysical Research: Space Physics, 125, \dodoi{10.1029/2019ja027084}

\bibitem[{Savani {et~al.}(2010)Savani, Owens, Rouillard, Forsyth, \& Davies}]{savani_2010_observational}
Savani, N., Owens, M.~J., Rouillard, A.~P., Forsyth, R.~J., \& Davies, J.~C. 2010, The Astrophysical Journal Letters, 714, L128, \dodoi{10.1088/2041-8205/714/1/l128}

\bibitem[{Schwenn(2006)}]{schwenn_2006_space}
Schwenn, R. 2006, Living Reviews in Solar Physics, 3, \dodoi{10.12942/lrsp-2006-2}

\bibitem[{Scolini {et~al.}(2021)Scolini, Dasso, Rodriguez, Zhukov, \& Poedts}]{scolini_2021_exploring}
Scolini, C., Dasso, S., Rodriguez, L., Zhukov, A.~N., \& Poedts, S. 2021, Astronomy and Astrophysics, 649, A69, \dodoi{10.1051/0004-6361/202040226}

\bibitem[{Scolini {et~al.}(2022)Scolini, Winslow, Lugaz, Salman, Davies, \& Galvin}]{scolini_2022_causes}
Scolini, C., Winslow, R.~M., Lugaz, N., {et~al.} 2022, The Astrophysical Journal, 927, 102, \dodoi{10.3847/1538-4357/ac3e60}

\bibitem[{Shiota \& Kataoka(2016)}]{shiota_2016_magnetohydrodynamic}
Shiota, D., \& Kataoka, R. 2016, Space Weather, 14, 56, \dodoi{10.1002/2015sw001308}

\bibitem[{Singh {et~al.}(2022)Singh, Kim, Pogorelov, \& Arge}]{singh_2022_ensemble}
Singh, T., Kim, T.~K., Pogorelov, N.~V., \& Arge, C.~N. 2022, The Astrophysical Journal, 933, 123, \dodoi{10.3847/1538-4357/ac73f3}

\bibitem[{Subramanian {et~al.}(2014)Subramanian, Arunbabu, Vourlidas, \& Mauriya}]{subramanian_2014_selfsimilar}
Subramanian, P., Arunbabu, K.~P., Vourlidas, A., \& Mauriya, A. 2014, The Astrophysical Journal, 790, 125, \dodoi{10.1088/0004-637x/790/2/125}

\bibitem[{Sudar {et~al.}(2022)Sudar, Vršnak, Dumbović, Temmer, \& Čalogović}]{sudar_2022_influence}
Sudar, D., Vršnak, B., Dumbović, M., Temmer, M., \& Čalogović, J. 2022, Astronomy and Astrophysics, 665, A142, \dodoi{10.1051/0004-6361/202244114}

\bibitem[{Temmer {et~al.}(2014)Temmer, Veronig, Peinhart, \& Vršnak}]{temmer_2014_asymmetry}
Temmer, M., Veronig, A., Peinhart, V., \& Vršnak, B. 2014, The Astrophysical Journal, 785, 85, \dodoi{10.1088/0004-637x/785/2/85}

\bibitem[{Temmer {et~al.}(2021)Temmer, Holzknecht, Dumbović, Vršnak, Sachdeva, Heinemann, Dissauer, Scolini, Asvestari, Veronig, \& Hofmeister}]{temmer_2021_deriving}
Temmer, M., Holzknecht, L., Dumbović, M., {et~al.} 2021, Journal of Geophysical Research: Space Physics, 126, \dodoi{10.1029/2020ja028380}

\bibitem[{Thernisien(2011)}]{thernisien_2011_implementation}
Thernisien, A. 2011, The Astrophysical Journal Supplement Series, 194, 33, \dodoi{10.1088/0067-0049/194/2/33}

\bibitem[{Vandas \& Romashets(2017)}]{mvandas_2017_magnetic}
Vandas, M., \& Romashets, E. 2017, Astronomy and Astrophysics, 608, A118, \dodoi{10.1051/0004-6361/201731412}

\bibitem[{Verbeke {et~al.}(2019)Verbeke, Pomoell, \& Poedts}]{verbeke_2019_the}
Verbeke, C., Pomoell, J., \& Poedts, S. 2019, Astronomy and Astrophysics, 627, A111, \dodoi{10.1051/0004-6361/201834702}

\bibitem[{Verbeke {et~al.}(2022)Verbeke, Schmieder, Démoulin, Dasso, Grison, Samara, Scolini, \& Poedts}]{verbeke_2022_overexpansion}
Verbeke, C., Schmieder, B., Démoulin, P., {et~al.} 2022, Advances in Space Research, 70, 1663, \dodoi{10.1016/j.asr.2022.06.013}

\bibitem[{Vourlidas {et~al.}(2010)Vourlidas, Howard, Esfandiari, Patsourakos, Yashiro, \& Michalek}]{vourlidas_2010_comprehensive}
Vourlidas, A., Howard, R.~A., Esfandiari, E., {et~al.} 2010, The Astrophysical Journal, 722, 1522, \dodoi{10.1088/0004-637x/722/2/1522}

\bibitem[{Vršnak {et~al.}(2008)Vršnak, Vrbanec, Čalogović, \& Žic}]{bojanvrnak_2008_the}
Vršnak, B., Vrbanec, D., Čalogović, J., \& Žic, T. 2008, Proceedings of the International Astronomical Union, 4, 271, \dodoi{10.1017/s1743921309029391}

\bibitem[{Vršnak {et~al.}(2013)Vršnak, Žic, Vrbanec, Temmer, Amerstorfer, Möstl, Veronig, Čalogović, Dumbović, Lulić, Moon, \& Shanmugaraju}]{bojanvrnak_2013_propagation}
Vršnak, B., Žic, T., Vrbanec, D., {et~al.} 2013, Journal of Geophysical Research, 285, 295, \dodoi{10.1007/s11207-012-0035-4}

\bibitem[{Vršnak {et~al.}(2010)Vršnak, Žic, Falkenberg, Möstl, Vennerstrøm, \& Vrbanec}]{bojanvrnak_2010_the}
Vršnak, B., Žic, T., Falkenberg, T.~V., {et~al.} 2010, Astronomy and Astrophysics, 512, A43, \dodoi{10.1051/0004-6361/200913482}

\bibitem[{Webb \& Howard(2012)}]{webb_2012_coronal}
Webb, D.~F., \& Howard, T.~A. 2012, Living Reviews in Solar Physics, 9, \dodoi{10.12942/lrsp-2012-3}

\bibitem[{Winslow {et~al.}(2021)Winslow, Scolini, Lugaz, \& Galvin}]{winslow_2021_the}
Winslow, R.~M., Scolini, C., Lugaz, N., \& Galvin, A.~B. 2021, The Astrophysical Journal, 916, 40, \dodoi{10.3847/1538-4357/ac0439}

\bibitem[{Xie(2004)}]{xie_2004_cone}
Xie, H. 2004, Journal of Geophysical Research, 109, \dodoi{10.1029/2003ja010226}

\bibitem[{Zhang {et~al.}(2004)Zhang, Dere, Howard, \& Vourlidas}]{zhang_2004_a}
Zhang, J., Dere, K.~P., Howard, R.~A., \& Vourlidas, A. 2004, The Astrophysical Journal, 604, 420, \dodoi{10.1086/381725}

\end{thebibliography}
\bibliographystyle{aasjournal}



\end{document}